\documentclass[preprint,12pt]{elsarticle}

\usepackage{amssymb}

\usepackage{amsthm}

\usepackage{graphicx}
\usepackage{amsmath}
\usepackage{comment}
\usepackage{paralist}
\usepackage[linesnumbered,ruled]{algorithm2e}
\usepackage{tikz}
\usepackage{soul}
\usetikzlibrary{automata, arrows.meta, positioning}

\usepackage[commandnameprefix=always]{changes}
\definechangesauthor[name={Ansuman}, color=violet]{AB}
\definechangesauthor[name={Kingshuk}, color=purple]{KC}
\definechangesauthor[name={Shibashis}, color=teal]{SG}

\newcommand{\NP}{{\sf NP}}
\newcommand{\NL}{{\sf NL}}
\newcommand{\PTIME}{{\sf PTIME}}
\newcommand{\EXPS}{{\sf EXPSPACE}}
\newcommand{\ins}{{\sf ins}}
\newcommand{\op}{{\sf op}}
\newcommand{\lsafa}{{\mathcal{L}_{\sf SAFA}}}
\newcommand{\ldsafa}{{\mathcal{L}_{\sf DSAFA}}}
\newcommand{\lkrfa}{{\mathcal{L}_{\sf KRFA}}}
\newcommand{\lcma}{{\mathcal{L}_{\sf CMA}}}
\newcommand{\lcca}{{\mathcal{L}_{\sf CCA}}}
\newcommand{\lhra}{{\mathcal{L}_{\sf HRA}}}
\newcommand{\cnt}{{\sf cnt}}

\newcommand{\regexp}{{\sf regexp}}
\newcommand{\projs}{{{\sf proj}_\Sigma}}
\newcommand{\projd}{{{\sf proj}_D}}

\newcommand{\zug}[1]{\langle #1 \rangle}

\newtheorem{example}{Example}
\newtheorem{definition}{Definition}
\newtheorem{theorem}{Theorem}
\newtheorem{lemma}{Lemma}
\newtheorem{corollary}{Corollary}

\begin{document}

\begin{frontmatter}



\title{Set Augmented Finite Automata over Infinite Alphabets}


\author[inst1]{Ansuman Banerjee}

\affiliation[inst1]{organization={Indian Statistical Institute},
            city={Kolkata},
            country={India}}

\author[inst2]{Kingshuk Chatterjee}
\author[inst3]{Shibashis Guha}

\affiliation[inst2]{organization={Government College of Engineering and Ceramic technology},
            city={Kolkata},
            country={India}}

\affiliation[inst3]{organization={Tata Institute of Fundamental Research},
            city={Mumbai},
            country={India}}

\begin{abstract}
 A data language is a set of finite words defined over an infinite alphabet. Data languages are used to express properties associated with data values (domain defined over a countably infinite set). In this paper, we introduce set augmented finite automata (SAFA), a new class of automata for expressing data languages. We investigate the decision problems, closure properties, and expressiveness of SAFA.
We also study the deterministic variant of these automata.
\end{abstract}




\end{frontmatter}


\section{Introduction}
 A data language is a set of data words that are concatenations of attribute, data-value pairs. While the set of attributes is finite, the values that these attributes hold often come from a countably infinite set (e.g. natural numbers). With
the availability of large-scale data in recent times, there is a need for methods for
modeling and analysis of data languages.
Thus, there is a demand for automated methods for recognizing attribute data relationships and languages defined on infinite alphabets. 
 A data word is a concatenation of a finite number of attribute, data-value pairs, i.e. a data word $w=(a_1,d_1)(a_2,d_2)...(a_{|w|},d_{|w|})$, where each $a_i$ belongs to a finite set and each $d_i$ belongs to a countably infinite set. We denote by $|w|$ the length of $w$. 
 This work introduces a new model for data languages on infinite alphabets. 
 $k$-register automata (finite automata with $k$ registers, each capable of storing one data value)~\cite{1} are finite automata with the ability to handle infinite alphabets. The nonemptiness and membership problems for register automata are both \NP-complete~\cite{4}. 
However, the language recognizability is somewhat restricted, since it uses only a finite number of registers to store the data values. 
Thus, register automata cannot accept many data languages, one such being $L_{{\sf fd}(a)}$\footnote{The article~\cite{14} contains a detailed review on infinite alphabet automata, we also follow the naming convention used in~\cite{14} to describe the languages.} which is a collection of data words where all data values associated with the attribute {\em $a$} have to be distinct. 
Pushdown versions of register automata on infinite alphabets using stacks have also been introduced in~\cite{2}. 
However, even with the introduction of a stack, these models cannot accept $L_{{\sf fd}(a)}$.
Data automata are introduced in~\cite{bojanczyk2006two}, and the emptiness problem is shown as nonelementary.
Further, class memory automata (CMA) and class counter automata (CCA) are introduced in~\cite{12} and ~\cite{13} respectively. 
While CMA and data automata are shown to be equivalent~\cite{12}, the set of languages accepted by CCA is a subset of the set of languages accepted by CMA. 
The nonemptiness problem for CCA is \EXPS-complete~\cite{13}, and the nonemptiness problem for CMA is interreducible to the reachability problem in Petri nets~\cite{14,bojanczyk2006two,Kosaraju82,Mayr84}, and hence Ackermann-complete~\cite{WL21}. 
Semilinear Data logic and semilinear data automata are introduced by Figuera et al.~\cite{figueira2022reasoning}. These can verify whether data values conform to one variable Presburger formulae. 
Semilinear data automata (SDA) and data automata are incomparable with respect to expressiveness. \\

\noindent
{\em Our Contribution:}
Our work introduces set augmented finite automata (SAFA), which are finite automaton models equipped with a finite number of finite sets for storing data values. Using these sets as auxiliary storage, SAFA is able to recognize many data languages mentioned in~\cite{14} including $L_{{\sf fd}(a)}$.
This paper has the following contributions. 
\begin{itemize}
  \item We present the formal definition of the SAFA model on infinite alphabets (Definition~\ref{safa_def}).
  
    \item We show that nonemptiness and membership are \NP-complete for SAFA. In addition, we show that the universality for SAFA is undecidable even for those with only one set.(Theorem~\ref{emptiness1},~\ref{member},~\ref{thm:univ}). 
    
    \item We study the closure properties of the SAFA model (Section~\ref{close}).  

    \item We also study the deterministic variant of SAFA and show that there are languages that necessarily need nondeterminism 
    to be accepted (Theorem~\ref{thm:dsafavsnsafa}). 
  
    \item We present a strict hierarchy of languages with respect to the number of sets associated with SAFA models (Theorem~\ref{thm:set_hier}).

    \item Finally, we compare the expressiveness of SAFA models with other automata models recognising data languages (see Section~\ref{sec:exp}). 
    
\end{itemize}

The elegance of SAFA is its simple structure, which is easy to implement. Moreover, the membership and nonemptiness problems are \NP-complete for our model.
On the one hand, this gives us an advantage over the models that use hash functions (CCA, CMA) with respect to the associated problem complexities. On the other hand, this puts our model in the same complexity class as $k$-register automata while having the ability to accept many data languages like $L_{{\sf fd}(a)}$ as described in~\cite{14} which $k$-register automata cannot accept.

\noindent
{\em Related work:} 
Register automata introduced by Kaminski et al.~\cite{1} 
use a finite number of registers to store data values; hence, they can only accept data languages in which membership depends on remembering properties of a finite number $k$ of data values where $k$ is bounded above by the number of registers in the register automata. A modification of register automata called guess register automata was introduced by Czerwi{\'n}ski et al.~\cite{czerwinski2021new} which allowed guessing of data values not yet seen which led to the universality problem even for one register to be undecidable.
Guess register automata is more expressive than register automata~\cite{czerwinski2021new} in terms of language acceptance. 
Since a guess register automaton also has a finite number of registers it cannot store all the data values it comes across in its input data word, thus cannot accept $L_{{\sf fd}(a)}$. 
A kind of a pumping lemma for languages accepted by one-register alternating finite-memory automata is in~\cite{DBLP:conf/stacs/Danieli26}.
An extension to finite register models is pushdown automata models for infinite alphabets \cite{2}. Cheng et al.~\cite{2} and Autebut et al.~\cite{3} both described context-free languages for infinite alphabets. The membership problem for the grammar proposed by Cheng et al. is decidable, unlike Autebut's. 
Sakamoto et al.\cite{4} showed that the membership problem is {\sf PTIME}-complete and nonemptiness is {\sf NP}-complete for deterministic register automata on infinite alphabets. 
Neven et al.\cite{5,6} discussed the expressiveness and decision problems such as emptiness, universality of register automata, tree register automata, and pebble automata on infinite alphabets and also established their relationship with logic. Tan et al.~\cite{15} introduced a weak 2-pebble automata model whose emptiness is decidable unlike pebble automata on infinite words~\cite{6} but with a significant reduction in acceptance capabilities with respect to pebble automata. 
Kaminski et al.\cite{7} also developed regular expressions over infinite alphabets. 
Choffrut et al.\cite{Choffrut1} defined finite automata on infinite alphabets where transitions between states are governed by first-order logic, i.e. if the corresponding data value of a data word satisfies the logical formula, then the transition is executed successfully. The paper~\cite{Choffrut1} was later extended by Iosif et al. to an alternating automata model \cite{9}, with the emptiness problem being undecidable. However, they proposed two efficient semi-algorithms for checking emptiness.  
Demri et al.~\cite{10} explored the relationship between linear temporal logic (LTL) over data words and register automata. 
Grumberg et al.~\cite{11} introduced variable automata over infinite alphabets where transitions are defined over alphabets and over variables. 
Data automata are introduced in~\cite{bojanczyk2006two}. 
Class automata~\cite{BL10} are strictly more expressive than data automata at the cost of undecidable emptiness checking. Two-variable logic on data words~\cite{bojanczyk2006two} was extended with guarded regular binary predicates in~\cite{DBLP:conf/icalp/GuhaMR26}. 
A new model of data automata namely set automata which is equivalent to class automata was also introduced in~\cite{DBLP:conf/icalp/GuhaMR26} and was shown to capture this logic.
While emptiness of set automata is undecidable, a restriction of this logic is also defined in~\cite{DBLP:conf/icalp/GuhaMR26} which leads to decidability through a translation to a restriction over set automata called ordered quasi-normal set automata which strictly includes the class of data automata~\cite{bojanczyk2006two}.
Class memory automata (CMA) were introduced in~\cite{12} and shown to be equivalent to data automata. The set of languages CMA accepts is a superset of Class counter automata (CCA)~\cite{13}, another infinite alphabet automata introduced by Manuel et al.  
Bollig~\cite{16} combined CMA and register automata and showed that local existential monadic second-order (MSO) logic could be converted to a class register automata in time polynomial in the size of the input automata. Figueira~\cite{17} discussed the properties of alternating register automata and restricted model variants where decidability is tractable.  Dassow and  Vaszil~\cite{18} introduced the {\em P}-automata model and established that these are equivalent to a restricted  version of register automata. History-register automata~\cite{grigore2016history} is a model introduced by Grigore et al. where the registers can remember a set of names, and the model can reset the registers. Bojanczyk et al.~\cite{bojanczyk2020single} introduced a single use register automata model where the content of a register gets erased after every access.   
Figueira et al.~\cite{figueira2022reasoning} proposed a parametric semilinear data logic (pSDL) for reasoning about data words with numeric data.
A detailed survey of existing finite automata models on infinite alphabets can be found in~\cite{14}. A detailed study of slightly infinite sets and automata can be found in~\cite{bojanczyk2019slightly}.

A preliminary version of this paper appeared in~\cite{BCG23}. The remainder of the paper is organized as follows. Section~\ref{sec3} discusses some preliminary concepts. Section~\ref{sec4} presents our SAFA model, while Section~\ref{sec5} discusses its decision problems and closure properties. Section~\ref{sec:exp} compares the expressiveness of SAFA with other models. Section~\ref{sec7} concludes the paper.

\section{Preliminaries} \label{sec3}

\noindent
Let $\mathbb{N}$ denote the set of natural numbers, and $[k]$ the set $\{1,\dots, k\}$ where $k>0$.
Let $\Sigma$ be a finite alphabet that denotes a finite set of attributes and let $D$ be a countably infinite set of attribute values.  
A \emph{data word} $w \in (\Sigma \times D)^*$ is a finite sequence of attribute, data value pairs; each attribute, data value pair is called a \emph{data item}, and $*$ denotes zero or more repetitions.
A data value is also known as an attribute value, i.e. the value associated with an attribute. A data word $w$ is of the form $(a_1, d_1) \cdots (a_{|w|}, d_{|w|})$\footnote{The parentheses are used to show the individual data elements and are not part of the alphabet.} where $a_1,\dots,a_{|w|}\in \Sigma$, $d_1,...,d_{|w|}\in D$, and $|w|$ denotes the length of $w$. An example of a data word $w$ on $\Sigma=\{a,b\}$ and $D=\mathbb{N}$ is: $(a,1)(a,2)(b,1)(b,5), (a,2)(a,5)(a,7)(a,100)$ with $|w|$ = 8. 
A {\em data language}  $L\ \subseteq(\Sigma \times D)^*$ is a set of data words. Some examples of data language with 
$\Sigma =\{a,b\}$, $D=\mathbb{N}$ that are used later are mentioned below. 
\begin{itemize}
  
    \item $L_{{\sf fd}(a)}$: the language of data words where the data values associated with attribute $a$ are all distinct.
    \item $L_{\forall \cnt=2}$: the  language of data words where all data values appear exactly twice.
    \item $L_{\exists \cnt \neq 2}$: the language of data words where there exists a data value $d$ which does not appear twice. $L_{\exists \cnt \neq 2}$ is the complement of $L_{\forall \cnt=2}$.
    \item $L_{a\exists b}$: the language of data words where the data values associated with attribute $a$ are those which have {\em already} appeared with the attribute $b$.
    
\end{itemize}

\noindent
For a word $w\in (\Sigma \times D)^*$, we denote by 
$\projs(w)$ and $\projd(w)$ the projection of $w$ on $\Sigma$ and $D$ respectively.
Let $L_{\projs (L)=\regexp(r)}$ be the set of all data words $w$ such that $\projs(w)\in L_{\regexp(r)}$ where $L_{\regexp(r)} \subseteq \Sigma^*$ is the set of all words over $\Sigma$ generated by the regular expression $r$.

For a word $w\in (\Sigma \times D)^*$ and sets $\theta_1 \subseteq \Sigma$, $\theta_2 \subseteq D$, 
we denote by 
${\sf sel}_{\theta_1\times \theta_2}(w)$ the subword $w'$ comprising of data items $(\alpha,\beta)$ where $\alpha \in \theta_1, \beta \in \theta_2$ appearing in the same order as they appear in $w$. For a data word $w=(a,5),(c,5)(b,5),(d,9)$, we have that ${\sf sel}_{\{c,d\}\times \mathbb{N}}(w)=(c,5)(d,9).$

\paragraph{Homomorphism on data words}
\noindent
A data word homomorphism is a function on data words that works by substituting each data item in a given data word. 
Formally if $h$ is a homomorphism from $(\Sigma \times D)^*$ to $(\Sigma' \times D')^*$ and $w=(a_1, d_1) \cdots (a_{|w|}, d_{|w|})$ where $a_1,\dots,a_{|w|}\in \Sigma$, $d_1,...,d_{|w|}\in D$ is a data word, then $h(w)=h((a_1,d_1))\cdots h((a_{|w|}, d_{|w|}))$, that is, $h$ is applied to each data item of $w$ and the result is concatenated, in order. 
Further, we can extend the notion of homomorphism to a language by applying it to each data word of the language. 
In other words, if $L$ is a language over the alphabet $(\Sigma \times D)$ and $h$ is a homomorphism over $(\Sigma \times D)^*$, then $h(L)=\{h(w)\:|\: w \in L\}$.

\paragraph{Inverse homomorphism on data words}
\noindent
Suppose $h$ is a homomorphism from $(\Sigma \times D)^*$ to $(\Sigma' \times D')^*$. Let $L$ be a language over $(\Sigma' \times D')$. Then $h^{-1}(L)=\{w\in (\Sigma\times D)^*|h(w)\in L\}$, i.e., the set of data words $w\in (\Sigma \times D)^*$ such that $h(w)\in L$.

\paragraph{$k$-register automata~\cite{1}}
\noindent
A $k$-register automaton is a tuple $(Q, \Sigma, \delta, \tau_0, U, q_0, F)$, where $Q$ is a finite set of states, $q_0 \in Q$ is an initial state
and $F \subseteq Q$ is a set of final states, $\tau_0$ is an initial register configuration given by $\tau_0 : [k] \rightarrow D \cup \{\bot\}$, where $D$ is a countably infinite set and $\bot $ denotes an uninitialized register, and $U$ is a partial update function: $(Q \times \Sigma) \rightarrow [k]$. 
The transition relation is $\delta \subseteq (Q \times \Sigma \times [k] \times Q)$. The registers initially contain distinct data values. Only the $\bot$ symbols can be present in more than one register in  $\tau_0$. The automaton works as follows.
Consider a register automaton $M$ in state $q\in Q$. Each of its registers $r_i$ holds datum $d_i$ where $0 \leq i \leq k$, $d_i\in D \cup \{\bot\}$. Let $M$ at some instance read the $j^{th}$ data element $(a_j,d_j)$ of the input word $w$ where $a_j\in \Sigma$, $d_j\in D$.
Two cases may arise. 
\begin{itemize}
\item Case 1: There exists an $i$ such that register $r_i$ contains $d_j$: In this case two situations may arise (i) if $(q,a_j,i,q')\in \delta$ for some $q'\in Q$ then the corresponding transition is executed or (ii) if $(q,a_j,i,q')\notin \delta$ for all $q'\in Q$ then the automaton stops without consuming the data element.
\item Case 2:  There exists no register $r_i$ such that $d_j=d_i$: In this case, for all $i$, we have $d_j\neq d_i$. We look at the partial update function $U$. If $U(q,a_j)$ is not defined, the automaton stops without consuming the data element. If $U(q,a_j)$ is defined, then $d_j$ is inserted in the register $U(q,a_j)$, and  the automaton executes the transition $(q,a_j,U(q,a_j),q')$ if $(q,a_j,U(q,a_j),q')\in \delta$, otherwise it halts if $(q,a_j,U(q,a_j),q')\notin \delta$. 
\end{itemize}

\noindent
The automaton $M$ accepts an input data word $w$ if it consumes the whole word and ends in a final state.

\paragraph{Class counting automata~\cite{13}}:
\noindent
A class counting automaton (a.k.a 1-bag CCA) is defined as a $5$-tuple $M=(Q,\Sigma ,\delta, q_0, F)$ where $Q$ is a finite set of states, $q_0 \in Q$ is an initial state, and $F\subseteq Q$ is a set of accepting states. 
A constraint $c$ is a pair $({\sf op},e)$, where ${\sf op}=\{<,>,=,\leq, \geq, \neq \}$, $e\in \mathbb{N}$. 
Let $C$ denote a collection of constraints.
The transition relation is $\delta\subseteq (Q\times \Sigma \times C \times Inst\times \mathbb{N}\times Q)$.
A bag is a finite set $\beta \subseteq (D\times \mathbb{N})$. 
Initially, $\beta(d)$ is set to $0$ for all data values $d \in D$. We store in the bag only some or all of those data values that we come across in the input data word, thus the bag is finite.
An element of ${\sf Inst} \times \mathbb{N}$ is an \emph{operation} where the set ${\sf Inst}=\{\uparrow^+,\downarrow\}$.
When transitioning, a CCA reads an attribute, data-value pair $(a,d)$, and checks if $\beta(d) \; {\sf op} \; e$ holds. If it holds, then (i) either $\beta(d)$ is incremented by $m$ if the operation is $(\uparrow^+, m)$, or (ii) $\beta(d)$ is reset to $m$ if the operation is $(\downarrow, m)$, and we go to the next state. 
A CCA accepts a data word $w$ if it is in a final state after consuming $w$.

A $k$-bag CCA has $k$ bags.
For a data value, constraint checking can be done on a subset of the bags.
 The bags can also be updated or reset independently. 
The set of transitions of a $k$-bag CCA is a subset of $(Q\times \Sigma \times C^k \times (Inst \times \mathbb{N})^k \times Q)$.
We denote by $\beta_i$, the $i^\text{th}$ bag.
It is shown in~\cite{13} that for every $k$-bag CCA that accepts a language $L$, there exists a $1$-bag CCA, which accepts the same language $L$. 
\paragraph{Class memory automata~\cite{12}}:
\noindent
A class memory automaton is a $6$-tuple $M=(Q,\Sigma ,\delta, q_0, F_{\ell},F_g)$ where $Q$ is a finite set of states, $q_0 \in Q$ is an initial state and $F_g\subseteq F_{\ell}\subseteq Q$ are a set of global and local accepting states respectively. The transition relation is $\delta\subseteq (Q\times \Sigma \times (Q\cup \{\bot \})\times Q)$.
The automaton keeps track of the last state where a data value $d$ is encountered. If a data value $d$ is not yet encountered, it is associated with $\bot$. Each transition of a CMA depends on the current state of the automaton and the state the automaton was in when the data value being read currently was last encountered. A data word $w$ is accepted if the automaton reaches a state $q\in F_g$ and the last state of all the data values encountered in $w$ are in $F_{\ell}$.
\paragraph{Semilinear Data automata~\cite{figueira2022reasoning}: }
\noindent
For a finite alphabet $\Sigma$, the Parikh
image of a word $w\in \Sigma^*$ is a function
$\Pi(w):\Sigma\rightarrow \mathbb{N}$ assigning to each $a\in \Sigma$ the number of appearances of $a$ in $w$.
Further, let $data(w)=\projd(w)$ and $lab(w)=\projs(w)$ denote the projection of $w$ on $\mathbb{N}$ and $\Sigma$, respectively.
Here $D=\mathbb{N}$.
Given a word $w\in \Sigma^*$ and a set $I \subseteq \{1, \dots , |w|\}$, let $w[I]$ denote the subword of $w$ given by the indices in $I$ (e.g., $w[\{1, \dots , |w|\}] = w,w[\phi] = \epsilon$). 
We use $w[i]$ instead of $w[\{i\}]$ for brevity.
A semilinear data automaton (SDA) over $\Sigma$ is a
pair $(T,S)$ where, for a finite alphabet $B$, we have:
\begin{enumerate}
\item $S$ is a semilinear set over $\mathbb{N}^B \times \mathbb{N}^B$.
\item $T \subseteq (\Sigma \times \mathbb{N})^* \times B^*$ is a length-preserving transducer, defined via a regular language $L_T \subseteq (\Sigma \times \psi \times B)^*$ for a finite set $\psi$ of quantifier-free Presburger formulas $\phi(x)$ with one free variable $x$ that is used to satisfy the property associated with the data value corresponding to an attribute, and the finite alphabet $B$. 
The symbol $T$ denotes the set of all pairs $(w,w') \in (\Sigma \times \mathbb{N})^* \times B^*$ such that there exists a sequence $\psi_1, \dots ,\psi_\ell$ of $\psi$-formulas where:
\begin{enumerate}
\item $|w| = |w'| = \ell$,
\item $data(w)[i]$ $\models \psi_i$ for all $i \in \ell$, that is $data(w)[i]$ satisfies $\psi_i$ and
\item $(lab(w)[1],\psi_1,w'[1]) \dots (lab(w)[\ell],\psi_\ell,w'[\ell])$ is in $L_T$. 
The alphabet $B$ is the output alphabet of $T$ and $\psi$ the Presburger
alphabet.
\end{enumerate}
\end{enumerate}
A word $w\in (\Sigma \times \mathbb{N})^*$ is accepted by an SDA if there exists some $w' \in B^
*$ such that
\begin{enumerate}
\item $(w,w') \in T$ , and 
\item for every $n\in \mathbb{N}$, we have that $(\Pi(w'[I_n]), \Pi(w'[\overline{I_n}])) \in S$,
where $I_n = \{1 \leq j \leq |w| : data(w)[j] =n\}$ and $\overline{I_n} = \{1, \dots , |w|\} \setminus I_n$.
\end{enumerate}
Given a set $S$, let $P(S)$ denote the set of all subsets of $S$, and $P_f(S)$ the set of all finite subsets of $S$. The symbol $\epsilon$ denotes the empty string.

\paragraph{History-register automata (HRA)~\cite{grigore2016history}:}
\noindent
An $(m,n)$-history-register automaton ($(m,n)\textit{-}HRA$) with $m$ sets and $n$ registers is a tuple $A=(Q,\Sigma, q_0,H_0,\delta,F)$ where:
\begin{itemize}
    \item $Q$ is a finite set of states, $q_0$ is an initial state, $F\subseteq Q$ is a set of accepting states.

    \item $H_0\in {\sf Asn}$ is an initial assignment where ${\sf Asn}=\{H:[m+n]\rightarrow P_f(D) \;\ \text{such that, we have} \;\ \forall i>m$, 
    also, $|H(i)|\leq 1\}$, is the set of all possible assignments of the sets and the registers such that each set has a finite number of data values and
 each register contains at most one data value.

    \item $\delta\subseteq Q\times \Sigma\cup \{\epsilon\} \times Lab \times Q$ is a transition relation where 
    $Lab=P([m+n])^2 \cup P([m+n]).$
\end{itemize}

\noindent
The transitions in HRA are of the form $q\xrightarrow[]{a,X,X'}q'$ and $q\xrightarrow[]{\epsilon, X}q'$, where $a\in \Sigma, X,X'\in P([m+n]), q,q'\in Q$. The transition $q\xrightarrow[]{a,X,X'}q'$ is enabled when the HRA $A$ in state $q$ reads the symbol $a$ and some data value $d$, and if the data value $d$ is present \emph{only} in each of the sets and registers in the set $X$, then the data value $d$ is removed from all the sets and registers in $X$ and is added to the sets and registers in $X'$. The HRA then goes to state $q'$. The transition $q\xrightarrow[]{\epsilon,X}q'$ denotes that the HRA $A$ in state $q$ does not read anything, and does not move its head from its current position but empties all the sets and registers in $X$. The HRA then goes to the state $q'$. 

A configuration of an $(m,n)\textit{-}HRA$ $A$ is a pair $(q,H)\in \widehat{Q}$, where $\widehat{Q}=Q\times {\sf Asn}$. An HRA accepts a data word $w$ if after completely consuming the data word $w$, the HRA $A$ is in an accepting state.

\paragraph{Nondeterministic finite automata~\cite{hopcroft2001introduction}: }
\noindent
A Nondeterministic Finite Automaton (NFA) is a mathematical model of computation that accepts or rejects finite sequences of symbols (also called a word) over an alphabet. 
Formally, an NFA is defined as a 5-tuple: $M = (Q, \Sigma, \delta, q_0, F)$ where $Q$ is a finite set of states, $\Sigma$ is a finite set of input symbols (the alphabet), $\delta : Q \times \Sigma \rightarrow 2^Q$ is the transition function, which maps a state and an input symbol to a set of possible next states,
where $2^Q$ denotes the power set of $Q$ (the set of all subsets of $Q$), $q_0$ is an initial state, and $F\subseteq Q$ is a set of final or accepting states. An NFA accepts an input string $w$ if and only if at least one path exists through the automaton, starting from the initial state, that leads to a final (accepting) state after consuming the entire input string. 
Size of an NFA $M=(Q,\Sigma ,q_0,F,\delta)$ is defined as $|M|=|Q|+\sum_{q\in Q, a\in \Sigma } |\delta(q,a)|$.

\paragraph{Deterministic finite automata~\cite{hopcroft2001introduction}: }
\noindent
A Deterministic Finite Automaton (DFA) is a mathematical model of computation that accepts or rejects finite sequences of symbols over an alphabet. 
Formally, a DFA is defined as a 5-tuple: $M = (Q, \Sigma, \delta, q_0, F)$ where $Q$ is a finite set of states, $\Sigma$ is a finite set of input symbols (the alphabet), $\delta : Q \times \Sigma \rightarrow Q$ is a transition function that maps a state and an input symbol to a next state (the transition function of DFA can be considered to be a special case of NFA where $|\delta(q,a)|=1, \text{for }q\in Q \text{ and }a\in \Sigma$), $q_0$ is an initial state and $F\subseteq Q$ is a set of final or accepting states. A string $w\in \Sigma^*$ is accepted by $M$ if and only if $\delta^*(q_0, w) \in F$, where $\delta^* : Q \times \Sigma^* \rightarrow Q$ is the extended transition function that maps a state and a string to the state reached after processing the entire string.

\section{Set augmented finite automata} \label{sec4}
\begin{definition}\label{safa_def}
A set augmented finite automaton (SAFA) is defined as a $6$-tuple $M = (Q,\Sigma \times D, q_0, F, H, \delta)$ where $Q$ is a finite set of states, $\Sigma$ is a finite alphabet, $D$ is a countably infinite set, $q_0 \in Q$ is an initial state, $F \subseteq Q$ is a set of final states, $H$ is a finite set of finite sets of data values. The transition relation is defined as: $\delta$ $\subseteq Q \times \Sigma \times C \times OP \times Q$ where $C= \{ p(h_i), !p(h_i) \ | \ h_i \in H\}$, $h_i$ denotes the $i^{th}$ set in $H$, and $OP = \{- ,\ \ins(h_i) \ | \  h_i \in H\}$. \qed
\end{definition}
We call a SAFA \emph{singleton} if $|H|=1$. The unary Boolean predicate $p(h_i)$ evaluates to true if the data value currently being read by the automaton is 
present in the $i^{th}$ set $h_i$. The predicate $!p(h_i)$ is true if the data value currently being read is {\em not} in $h_i$. Further, $OP$ denotes a set of operations that a SAFA can execute on reading a symbol; the operation $\ins(h_i)$ inserts the data value currently being read by the automaton into the set $h_i$, while $-$ denotes no such insertion is done. 
For any entry not in $\delta$, we assume the transition is absent. 

For a SAFA $M$, we define a \emph{configuration} $(q,h) \in Q \times 2^{D^H}$ as follows: $q\in Q$ is a state of the automaton, $h=\langle h_1,...h_{|H|} \rangle$ where each $h_i$ for $i \in [|H|]$ is a finite subset of $D$, and $h$ denotes the content of the sets in $H$. 
A \emph{run} $\rho$ of $M$ on an input $w=(a_1,d_1) \cdots (a_{|w|}, d_{|w|})$ is a sequence $(q_0,h^0), \dots, (q_{|w|},h^{|w|})$, where $h^j=\zug{h^j_1, \dots, h^j_{|H|}}$, and $h^j_i$ for $1 \le i \le |H|$ is the content of the set $h_i$ after reading the prefix $(a_1,d_1) \cdots (a_j, d_j)$ for $1 \le j \le |w|$.
A configuration $(q_{j+1},h^{j+1})$ succeeds a configuration $(q_j,h^j)$ if there is a transition $(q_j,a_{j+1},\alpha,\op,q_{j+1})$ where
 \begin{enumerate}[(i)]
     \item for $\alpha=p(h_i)$, we have the data value $d_{j+1}\in {h_i}^j$.
     \item for $\alpha \ = \ !p(h_i)$, we have  the data value $d_{j+1}\notin {h_i}^j$.
 \end{enumerate}
  The execution of the operation $\op \in OP$ defines the content of the sets of data values in $h^{j+1}$ in terms of $h^{j}$.
  If $\op$ is $-$, then $h^{j+1}=h^{j}$.  
If $\op$ is \ins$(h_i)$, then ${h_l}^{j+1}={h_l}^j$ for all $ h_l\in H \setminus \{h_i\}$, and ${h_i}^{j+1}={h_i}^j\cup\{d_{j+1}\}$. 
If the run consumes the whole word $w$, and $q_{|w|} \in F$, then the run is \emph{accepting} and rejecting otherwise. 
The SAFA $M$ accepts a word $w$
if it has an accepting run. 
The language $L(M)$ recognised by $M$ consists of all words accepted by $M$.  We denote by $|\rho|$ the 
length of the run, which equals the number of transitions taken. Note that for the run $\rho$ on an 
input word $w$, we have that $|w|$ = $|\rho|$.

\begin{definition}
 A SAFA $M = (Q,\Sigma, q_0, F, H, \delta)$ is \emph{deterministic} (we call it DSAFA) if for every $q \in Q$, $a \in \Sigma$, there exists a unique $h_i \in H$ such that there can be at most one transition of the form $(q,a,p(h_i), \op, q')$ and at most one transition of the form $(q,a,!p(h_i), \op', q'')$ for $\op, \op' \in OP$, $q',q'' \in Q$. 
In other words, for every $q \in Q$ and $a\in \Sigma$ if there is a transition $(q,a,\alpha,\op,q')$, where $q'\in Q$, $\op\in OP$, $\alpha\in \{p(h_i),!p(h_i)\}$, $h_i\in H$, then there cannot be any transition of the form $(q,a,p(h_l),\op',q'')$ and $(q,a,!p(h_l),\op',q'')$, where $q''\in Q$, $h_l\neq h_i$, $h_l\in H$, $\op'\in OP$. However for a particular $h_i$, both $(q,a,p(h_i),op,q')$ and $(q,a,!p(h_i),op',q'')$ can be there. \qed
\end{definition}
  
\noindent 

Let $\lsafa$ and $\ldsafa$ denote the class of languages accepted by nondeterministic SAFA and deterministic SAFA respectively. We illustrate the SAFA model with some instances of data languages. 

\begin{figure} [t]
\centering
\begin{tikzpicture} [node distance = 3cm, on grid, auto]

\node (q0) [state, initial, accepting, initial text = {}] {$q_0$};
\node (q1) [state, right = of q0] {$q_1$};

\path [-stealth, thick]
	(q0) edge [loop below]  node [right] {{\footnotesize $(a,!p(h_1),\ins(h_1))$}}    (q0)
	(q0) edge [loop above]  node [right] {{\footnotesize $(b,p(h_1),-), (b,!p(h_1),-) $}}    (q0)
    (q0) edge  node[above] {{\footnotesize $(a,p(h_1),-)$}}    (q1);
\end{tikzpicture}
\begin{centering}
\caption{SAFA for $L_{{\sf fd}(a)}$} \label{fig1}
\end{centering}
\end{figure}

\begin{example}\label{eg1}
The language $L_{{\sf fd}(a)}$ 
can be accepted by the DSAFA $M=(Q,\Sigma \times D,q_0,F,H,\delta)$  in Figure~\ref{fig1}. Here, $Q=\{q_0,q_1\}$, $\Sigma=\{a,b\}$, $D$ is any countably infinite set, $F=\{q_0\}$, $H=\{h_1\}$, the transition relation $\delta$ consists of the transitions shown in Figure~\ref{fig1}.
 \noindent
The automaton $M$ works as follows. The set $h_1$ is used to store the data values encountered in the input word associated with $a$. At $q_0$, if $M$ reads $b$, it remains in $q_0$ without modifying $H$. At $q_0$ when the automaton reads $a$, it checks whether the corresponding data value is present in $h_1$. If present, it indicates that it has already encountered this data value with attribute $a$ before, and the automaton moves to state $q_1$ which is a dead state and the input word is rejected. If the data value is not present in $h_1$, it implies that this data has not been encountered before with attribute $a$. The SAFA $M$ then remains in $q_0$ and inserts the data value into $h_1$. Thus, if the SAFA $M$ does not encounter duplicate data values with respect to attribute $a$ in the input, it remains in $q_0$. After consuming the entire word, it is accepted. \qed
\end{example}

\begin{example}\label{eg4}
Consider the data language $L_{\exists \cnt \neq 2:}$ over the alphabet 
$\Sigma=\{a\}$. A nonempty word $w$ is in the language if there exists a data value that appears $n$ times 
in $w$ with $n$ $\neq$ 2.
The nondeterministic SAFA shown in Figure~\ref{fig3} can accept the above language. 
At state $q_0$, the automaton nondeterministically guesses the data value that does not appear exactly twice and stores it in set $h_2$ and goes to state $q_1$. The set $h_1$ is needed to ensure that when a data value is nondeterministically decided to be the one that does not appear twice, it is the first instance of the data value in the data word. The automaton remains in state $q_1$ if the count of the data value 
is $1$ or moves to $q_3$ (via $q_2$) and remains there if the count of the data value is greater than $2$. In both cases, it accepts the input word once it is consumed entirely. If the guess is incorrect, the data value appears twice; then it is in the nonaccepting state $q_2$ after consuming the input word. Thus, if a data word has all its data values that appear exactly twice, all the runs end in either $q_0$ or in $q_2$, and the input is rejected. \qed

\end{example}

\begin{figure} [t]
\centering
\begin{tikzpicture} [node distance = 3.5cm, on grid, auto]

\node (q0) [state, initial, initial text = {}] {$q_0$};
\node (q1) [state, accepting, right = of q0] {$q_1$};
\node (q2) [state, right = of q1] {$q_2$};
\node (q3) [state, accepting, right = of q2] {$q_3$};

\path [-stealth, thick]
    (q3) edge [loop below]   node [below left] {{\footnotesize $(a,!p(h_1),-), (a,p(h_1),-)$}}    (q3)
	(q0) edge [loop above]  node [above right] {{\footnotesize $(a,p(h_1),\ins(h_1)), (a,!p(h_1),\ins(h_1))$}}    (q0)
    (q0) edge   node[below]         {{\footnotesize $(a,!p(h_1),\ins(h_2))$}}    (q1)
    (q1) edge [loop below]  node [below right] {{\footnotesize $(a,!p(h_2),-)$}}    (q1)
    (q1) edge  node[above ]         {{\footnotesize $(a,p(h_2),-)$}}    (q2)
    (q2) edge [loop above]  node [above right] {{\footnotesize $(a,!p(h_2),-)$}}    (q2)
    (q2) edge   node[above]         {{\footnotesize $(a,p(h_2),-)$}}    (q3);
    
\end{tikzpicture}

\caption{SAFA for $L_{\exists \cnt \neq 2}$} \label{fig3}
\end{figure}

The number of sets in the SAFA model has an important role in the languages that can be accepted. The following theorem establishes a hierarchy of accepted languages by SAFA  based on the cardinality of $H$.

\begin{lemma}\label {lem:set_hier}
    There exists a SAFA with \(k+1\) sets which can accept the language $L=\{ww\:|\:w\in(\{a\} \times D)^{k+1}\}$. 
    No SAFA with \(k\) sets can accept the language \(L\).
\end{lemma}
\begin{proof}
We first show that there exists a SAFA with \(k+1\) sets that can accept the language $L$.
Consider a deterministic SAFA \(M = (Q,\{a\}\times D,q_0,F,H,\delta)\), where \(Q = \{q_0, q_1, \dots, q_{2k+2}\}\), \(F = \{q_{2k+2}\}, \textrm{and} \ H = \{h_1, \dots, h_{k+1}\}\) is a set of \(k+1\) sets.
The SAFA \(M\) has \(2(k+1)\) transitions in sequence where the first \(k+1\) transitions are of the form \((a, !p(h_i), \ins(h_i))\) for \(i \le k+1\).
The next \(k+1\) transitions are of the form \((a, p(h_i),-)\) for \(i \le k+1\).
It is straightforward to see that this SAFA accepts the language \(L\).

Now, we show that no SAFA with \(k\) sets can accept \(L\).
Suppose, for contradiction, there exists a SAFA \(M\) with \(k\) sets, which accepts \(L\).
Consider the word \(w=(a,d_1)(a,d_2),\dots (a,d_{k+1})(a,d_1)(a,d_2),\dots (a,d_{k+1})\) where all the data values \(d_1, \dots, d_{k+1}\) are different.
The word \(w\) is in \(L\) and hence accepted by \(M\).
Thus, there is an accepting run of length \(2(k+1)\); that is, the run consists of \(2(k+1)\) transitions.
The last \(k+1\) transitions are of the form \((a,p(h_i), \ins(h_j))\) or \((a,p(h_i),-)\) since any data value \(d_i\) in the second half of the word must have appeared exactly \(k+1\) positions before.
During the first occurrence of the data value \(d_i\), the transition reading $d_i$ stores it in a set so that during its second occurrence in the second half of \(w\), it can be checked that indeed the data value appeared earlier in the first half of \(w\).
Since each half of \(w\) has length \(k+1\) and \(M\) has \(k\) sets, it follows from the pigeonhole principle that there exist two data values \(d_i\) and \(d_j\) with \(i,j \le k+1\) such that both \(d_i\) and \(d_j\) are stored in the same set, say \(h\), while executing the first \(k+1\) transitions.
Thus, in order to accept \(w\), the transitions at positions \(k+1+i\) and \(k+1+j\) should be of the form \((a, p(h), \op_i)\) or  \((a, p(h), \op_j)\), where \(\op_i, \op_j \in \{\ins(h_\ell), \ins(h_m), -\}\) for \(\ell, m \in [k]\).

Let us assume without loss of generality, \(i < j\). Consider the word \(w'\) where the positions of \(d_i\) and \(d_j\) are interchanged in the second half, i.e., \(w' = (a,d_1) \dots (a,d_i) \dots (a,d_j) \dots (a, d_{k+1}) (a,d_1) \dots (a,d_j) \dots (a,d_i) \dots (a, d_{k+1})\).

While \(w' \notin L\), the SAFA \(M\) still accepts \(w'\).
Hence the result.
\end{proof}

\noindent
Let $\lsafa_{(|H|=k)}$ be the set of languages accepted by SAFA with $|H|=k$.
Note that every SAFA $M=(Q,\Sigma\times D,q_0,F,H,\delta)$ with $|H|=k$ can be simulated by a SAFA $M'=(Q',\Sigma\times D,q_0,F',H',\delta')$ with $|H'|=\ell$ and $\ell>k$ by using $\ell-k$ dummy sets that are never used in execution of $M'$. 
Thus, we have the following theorem as a corollary of Lemma~\ref{lem:set_hier}.

\begin{theorem}\label {thm:set_hier}
$\lsafa_{(|H|=k)} \subsetneq \lsafa_{(|H|=k+1)}$ for \(k \in \mathbb{N}\).
\end{theorem}

\noindent
Theorem~\ref{thm:set_hier} states that there is a strict hierarchy in terms of accepting capabilities of SAFA with respect to the number of sets it has.

\section{Decision problems and closure properties} \label{sec5}
In this section, we study the nonemptiness, membership problems and closure properties of SAFA.
\subsection{Nonemptiness and membership}

We study the nonemptiness and the membership problems of SAFA and show that both are $\NP$-complete. 
Given a SAFA $M$ and an input word $w$, the \emph{membership problem is to check if $w \in L(M)$.}
Given a SAFA $M$, the \emph{nonemptiness problem} is to check if $L(M) \neq \emptyset$.
We start with the nonemptiness problem. To show the NP-membership, we first show that there exists a small run if the language accepted by a given SAFA is nonempty.

\begin{lemma}\label{lem:pimin}
    Every SAFA $M=(Q,\Sigma \times D,q_0,F,H,\delta)$ with $L(M)\neq \emptyset$ has a data word in $L(M)$ with an accepting run $\rho$ such that $|\rho|$ $\leq$ $|Q| \cdot (|H|+2) - 1$.
\end{lemma}
\begin{proof}
    We prove by contradiction. 
    Assume that $L(M) \neq \emptyset$ and that for every $w \in L(M)$, for all accepting runs $\rho$ of $w$, we have that $|\rho| > |Q| \cdot (|H|+2)-1 = |Q| \cdot (|H|+1) + |Q|-1$.
    We define an indicator function $I_H$ which maps $H$ to $\{0,1\}^{H}$, where $0$ corresponding to a set $h \in H$ denotes that $h$ is empty while $1$ denotes that $h$ is nonempty. 
Since $|\rho| \geq |Q| \cdot (|H|+2)$, the run $\rho$ can be divided into $|H|+2$ segments, each of length $|Q|$, that is, each segment is an infix over $|Q|$ transitions.
    By the pigeonhole principle, there exists a state $q \in Q$ that is visited more than once in each segment.
    Recall that in a SAFA, once a set has some element in it, it cannot become empty later as elements cannot be removed.
    Thus, since there are $|H|+2$ such segments, again by pigeonhole principle, there exists a segment and a state $q' \in Q$ such that $q'$ is visited more than once in this segment and $I_H$ does not change over the infix of the run between the two successive visits of $q'$.
    Let $y$ be this infix over which $I_H$ does not change.
    We now note that the sequence of transitions reading this infix $y$ makes a loop over $q'$, and thus $y$ can be removed from $w$, resulting in a word $w'$ and the corresponding run is $\rho'$ such that $|\rho'| < |\rho|$ and $\rho'$ is accepting.
    Now, there can be two cases if the suffix of $\rho$ following reading $y$ in $w$ has a transition $t$ with $p(h_i)$ for some set $h_i \in H$, and it reads a data value $d$.
    \begin{enumerate}
    \item It may happen that the data value $d$ was inserted along the infix $y$.
    Since $I_H$ does not change while reading the infix $y$, it implies that the set $h_i$ was nonempty even before the infix $y$ was read. 
    Let a data value $d'$ be inserted into $h_i$ while reading the prefix before $y$.
    Then $w'$ may be modified to $w''$ so that the suffix following $y$ in $w''$ reads the data value $d'$ instead of $d$.
    Let the run corresponding to $w''$ be $\rho''$ that follows the same sequence of states as $\rho'$.
    Note that $|\rho''| = |\rho'| < |\rho|$, and that $\rho''$ is an accepting run.
    \item If while reading $w$, the transition $t$ reads a data value $d$ that was inserted while reading a prefix appearing before $y$, then $w'$ does not need to be modified, and we thus have the accepting run $\rho'$.
    \end{enumerate}
    Since $\rho$ is an arbitrary accepting run of length $|Q| \cdot (|H|+2)$ or more, starting from $\rho$, we can remove infixes repeatedly and modify it as mentioned above until we reach an accepting run of length strictly smaller than $|Q| \cdot (|H|+2)$ without affecting acceptance, and hence the contradiction.
\end{proof}

\noindent
Using Lemma~\ref{lem:pimin} we get the following.

\begin{lemma}\label{safa_np}
Nonemptiness problem for SAFA is in NP.
\end{lemma}
\begin{proof}
Consider a SAFA $M$ with $L(M) \neq \emptyset$. By Lemma~\ref{lem:pimin}, a Turing machine can nondeterministically guess an accepting run of polynomial length, hence the result.
\end{proof}

\noindent
We now show that the nonemptiness problem is NP-hard even for deterministic acyclic SAFA over an alphabet of size $3$. Example~\ref{npheg} describes our construction.
We show a reduction from 3SAT.

\begin{example}\label{npheg}
For a 3CNF formula $\phi=(x_1 \vee \overline{x_2}\vee {x_3}) \wedge (x_1 \vee x_2\vee x_3)$, 
the corresponding SAFA $M=(Q,\Sigma \times D,q_0,F,H,\delta)$ is shown in Figure~\ref{nphf}. We denote by $A(a,x_i)$ the transition $(a,!p(h_{x_i}),\ins(h_{x_i}))$ and  by $T(a,x_i)$ the transition $(a,p(h_{x_i}),-)$. Here
$Q=\{q_0,q_{x_1},q_{x_2},q_{x_3},q_{C_1}, q_{C_2}\}$, 
$\Sigma=\{a_1,a_2,a_3\}$, $D=\mathbb{N}$, 
$H=\{h_{x_1},h_{\overline{x_1}}, h_{x_2},h_{\overline{x_2}},h_{x_3},h_{\overline{x_3}}\}$, $F=\{q_{C_2}\}$. In particular, if there are $\ell$ variables in the formula, then $|H|=2\ell$. The $A$ transitions correspond to setting the Boolean variables while the $T$ transitions correspond to checking the clauses. \qed

\end{example}

\begin{figure}[t]
\centering
\begin{tikzpicture} [node distance = 2.5cm, on grid, auto]

\node (q0) [state, initial,  initial text = {}] {$q_0$};
\node (q1) [state, right = of q0] {$q_{x_1}$};
\node (q2) [state, right = of q1] {$q_{x_2}$};
\node (q3) [state, right = of q2] {$q_{x_3}$};
\node (q4) [state, right = of q3] {$q_{c_1}$};
\node (q5) [state, accepting, right = of q4] {$q_{c_2}$};
\path [-stealth, thick]
	(q0) edge [bend left, above]   node[above] 
     {{\footnotesize $A(a_1,x_1)$}}    (q1);
 
  \path [-stealth, thick]
	(q0) edge [bend right, below]   node[below] 
     {{\footnotesize $A(a_2,\overline{x_1})$}}    (q1);  
     
     \path [-stealth, thick]
	(q1) edge [bend left, above]   node[above] 
     {$A(a_1,x_2)$}  (q2);
 
  \path [-stealth, thick]
	(q1) edge [bend right, below]   node[below] 
     {{\footnotesize $A(a_2,\overline{x_2})$}}   (q2); 
     
     \path [-stealth, thick]
	(q2) edge [bend left, above]   node[above] 
     {{\footnotesize $A(a_1,x_3)$}}    (q3);
 
   \path [-stealth, thick]
	(q2) edge [bend right, below]   node[below] 
     {{\footnotesize $A(a_2,\overline{x_3})$}}   (q3); 

      \path [-stealth, thick]
	(q3) edge [bend left, above]   node[above] 
     {{\footnotesize $T(a_1,x_1)$}}    (q4);
 \path [-stealth, thick]
	(q3) edge  node[below] 
     {{\footnotesize $T(a_2,\overline{x_2})$}}   (q4);
   
   \path [-stealth, thick]
	(q3) edge [bend right=50, below]   node[below] 
      {{\footnotesize $T(a_3,x_3)$}}    (q4);

      \path [-stealth, thick]
	(q4) edge [bend left, above]   node[above] 
      {{\footnotesize $T(a_1,x_1)$}}    (q5);
 \path [-stealth, thick]
	(q4) edge  node[below] 
      {{\footnotesize $T(a_2,x_2)$}}    (q5);
   
   \path [-stealth, thick]
	(q4) edge [bend right=50, below]   node[below] 
      {{\footnotesize $T(a_3,x_3)$}}    (q5);

\node[draw,text width=4cm] at (2,-2) {{\footnotesize $A(a,x_i)=(a,!p(h_{x_i}),\ins(h_{x_i}))$}, {\footnotesize $T(a,x_i)=(a,p(h_{x_i}),-)$}}
;
	
\end{tikzpicture}
\caption{The SAFA $M$ corresponding to a 3CNF formula $\phi$} \label{nphf}
\end{figure}

\begin{lemma}\label{safa_hard}
The nonemptiness problem is \NP-hard for deterministic acyclic SAFA over an alphabet of size $3$.
\end{lemma}
\begin{proof}
    Let $\psi=C_1\wedge C_2\dots \wedge C_m$ be a CNF formula. Let $Var$ be the set of variables in $\psi$ and $C=\{C_1,C_2,\dots C_m\}$ be the set of clauses. We construct the SAFA with $M=(Q,\Sigma \times D,q_0,F,H,\delta)$ where $Q=\{q_0\}\bigcup \{(q_x)_{x\in Var}\}\bigcup \{(q_c)_{c\in C}\}$, $F=\{q_{C_m}\}$.
    Let there be $n$ variables $x_1,x_2,\dots x_n$ in $\psi$. The transitions in $M$ are as follows:

    We denote by $A(a,x_i)$ the transition $(a,!p(h_{x_i}),\ins(h_{x_i}))$ and by $T(a,x_i)$ the transition $(a,p(h_{x_i}),-)$ respectively.

    From  $q_0$ to $q_{x_1}$, there are two transitions labeled $A(a_1,x_1)$ and $A(a_2,\overline {x_1})$.

    For $i\in [n-1]$ from $q_{x_i}$ to $q_{x_{i+1}}$, there are two transitions labelled $A(a_1,x_{i+1})$ and $A(a_2,\overline {x_{i+1}})$.

    Now, we describe the transitions from $q_{x_n}$ to $q_{C_1}$. Let $C_1$ be of the form  $l_i\vee l_j\vee l_k$ where $i,j,k\in [n]$, and $l_p\in \{x_p,\overline{x_p}\}$ for $p\in \{i,j,k\}.$ 

    From $q_{x_n}$, we have transitions $T(a_1,l_i)$, $T(a_2,l_j)$\textcolor{red}{,} and $T(a_3,l_k)$ to $q_{C_1}$. Similarly, if $C_r$ is of the form $l_i\vee l_j\vee l_k$ where $i,j,k\in [n]$ and $r\in \{2,\dots m\}$, then we have transitions $T(a_1,l_i)$, $T(a_2,l_j)$ and $T(a_3,l_k)$ from $q_{C_{r-1}}$ to $q_{C_r}$.

    The set $F$ of final states  only consists of a single final state $q_{C_m}$.

    A run from $q_0$ to $q_{C_m}$ consists of $n+m$ transitions where the first $n$ transitions set values to the  variables $x_1$ to $x_n$ while the last $m$  transitions  check if the $m$ clauses are satisfied.

    It is easy to see that if the 3CNF  formula  $\psi$ is satisfiable, then there exists a  run in $M$ that is accepting. On the other hand, if $\psi$ is  not satisfiable, then for every assignment, there exists a clause  $C_i$ that cannot be satisfied. If $i=1$, none of the transitions from $x_n$ to $C_1$ can be taken.  Additionally, if $i>1$, none of the transitions from $C_{i-1}$ to $C_i$ can be taken. Therefore, there is no accepting run, and hence $L(M)=\phi$.   
\end{proof}

From Lemma~\ref{safa_np} and Lemma~\ref{safa_hard}, we have the following.
\begin{theorem} \label{emptiness1}
The nonemptiness problem for SAFA is \NP-complete. 
\end{theorem}
\noindent
\noindent 
We now show that for singleton SAFA, nonemptiness is \NL-complete.
Towards this, we first show the following lemma.
\begin{lemma} \label{lem:SAFAtoNFA}
The nonemptiness problem for singleton SAFA is reducible to the nonemptiness of a nondeterministic finite automaton (NFA) in \PTIME.
\end{lemma}
\begin{proof}
We begin with the following observations on SAFA transitions for a run on an input word. The idea is to see how we 
can construct a word accepted by a given SAFA that takes it from the initial state to a final state following the transition 
rules. Let $h_1$ be the set in a singleton SAFA. Note that transitions with $!p(h_1)$ can always be satisfied since we have an infinite number of data values. We can always introduce a new data value with an attribute so that it is not in the set $h_1$. However, transitions with $p(h_1)$ should only be executed if there exists an $\ins(h_1)$ somewhere earlier on the path before reaching the transition with $p(h_1)$. Hence, given a SAFA, it is not enough to only look for simple paths from the initial state to a final state. For  example, in Figure \ref{f4}, the only simple path is $q_0 \rightarrow q_f$ with transition $(a,p(h_1),-)$. Since the simple path does not contain any transition having $\ins(h_1)$ prior to the transition containing $p(h_1)$, no word is accepted by the automaton along this simple path. However, the automaton accepts the string $(a,d_1)(a,d_1)$. Thus, when checking for the emptiness of a SAFA, we cannot just restrict our analysis to simple paths. 
Thus, the language accepted by a SAFA is nonempty if and only if there exists a sequence of transitions, that takes it from an initial state to a final state with the added condition that if the sequence of transitions contains a $p(h_1)$, then there must exist a corresponding $\ins(h_1)$ prior to the $p(h_i)$ in the sequence of transitions. 

\begin{figure}[t]
\begin{minipage}{\textwidth}
\begin{minipage}{0.4\textwidth} 
\centering
\begin{tikzpicture} [node distance = 3cm, on grid, auto]

\node (q0) [state, initial, initial text = {}] {$q_0$};
\node (q1) [state, accepting, right = of q0] {$q_f$};

\path [-stealth, thick]
    (q0) edge  node[above]         {{\footnotesize $(a,p(h_1),-)$}}    (q1)
    (q0) edge [loop below]  node [right] {{\footnotesize $(a,!p(h_1),\ins(p(h_1)) $}}    (q0);
\end{tikzpicture}
\caption{A simple SAFA $M$}\label{f4}
\end{minipage}
\hfill
\begin{minipage}{0.6\textwidth} 
\centering
\begin{tikzpicture} [node distance = 3cm, on grid, auto]

\node (q0) [state, initial, initial text = {}] {$q_0$};
\node (q1) [state, accepting, right = of q0] {$q_f$};

\path [-stealth, thick]
    (q0) edge   node[above]         {{\footnotesize$(a,p(h_1),-)$}}    (q1)
    (q0) edge [loop below]  node [right] {{\footnotesize $(a,!p(h_1),\ins(p(h_1))$}}    (q0);

\end{tikzpicture}
\caption{The NFA $M_1$ corresponding to SAFA $M$} \label{f5}
\end{minipage}
\vspace{1pt} 
\begin{minipage}{\textwidth} 
\centering
\begin{tikzpicture} [node distance = 4cm, on grid, auto]

\node (q0) [state, initial, accepting, initial text = {}] {$q_0$};
\node (q1) [state, accepting, right = of q0] {$q_1$};

\path [-stealth, thick]
    (q0) edge   node[above]         {{\footnotesize $(a,!p(h_1),\ins(h_1))$,}}   (q1)
    (q0) edge [loop above]  node [left] {{\footnotesize$(a,!p(h_1),-)$ } }   (q0)
  (q1) edge [loop below]  node [below] {{\footnotesize $(a,\alpha,op)$,$\alpha \in \{p(h_1),!p(h_1)\}$, $op\in \{-,\ins(h_1)\}$}}    (q1);
    
\end{tikzpicture}
\caption{The NFA $M_2$ corresponding to SAFA M}\label{f6}
\end{minipage}
\end{minipage}
\vspace{1pt} 
  \begin{minipage}{\textwidth} 
  \centering
\begin{tikzpicture} [node distance = 5cm, on grid, auto]

\node (q0) [state, initial, initial text = {}] {$q_0q_0$};
\node (q1) [state, right = of q0] {$q_0q_1$};
\node (q2) [state, accepting, right = of q1] {$q_fq_1$};

\path [-stealth, thick]
   
    (q0) edge   node[above ]         {{\footnotesize $(a,!p(h_1),\ins(h_1))$}}    (q1)
    (q1) edge [loop above]  node [right] {$(a,!p(h_1),\ins(h_1))$}    (q1)
    (q1) edge   node[above]         {{\footnotesize $(a,p(h_1),-)$}}    (q2);
\end{tikzpicture}
\caption{NFA $M_3$ : Synchronous product of NFA $M_1$ and NFA $M_2$ for SAFA M}\label{f7}
\end{minipage}
\end{figure}

Given a singleton SAFA $M=(Q,\Sigma \times D,q_0,F,H,\delta)$, $H=\{h_1\}$, we check for nonemptiness by constructing three nondeterministic finite automata (NFA). The first NFA $M_1$ accepts all those sequences of transitions which {can} take the SAFA $M$ from its initial state to any of its final states. However, $M_1$ does not check if the sequence of transitions on a path from the initial state to a final state of $M$ is valid; that is, a transition with  $p(h_1)$ is preceded by a transition with $\ins(h_1)$ on the path.
The second NFA $M_2$ accepts all possible sequences of transitions that have $\ins(h_1)$ prior to encountering a $p(h_1)$. We construct the synchronous product~\cite{22} of $M_1$ and $M_2$ that gives us another NFA  $M_3$. We check for the emptiness of $M_3$. If $M_3$ is not empty (final state is reachable), we can conclude that there exists at least one sequence of transitions that takes the SAFA $M$ from the initial state to a final state and every $p(h_1)$ encountered on that sequence of transitions has an $\ins(h_1)$ prior to it. Therefore, the SAFA $M$ is not empty. If $M_3$ is empty, it indicates that there exists no such sequence of transitions.  
Thus, if $M_3$ is empty, we can conclude that the SAFA $M$ is also empty, and no data word is accepted 
by the SAFA. We give a formal construction below.

Given a SAFA $M=(Q,\Sigma \times D,q_0,F,H,\delta)$, $H=\{h_1\}$ we construct NFA $M_1=(Q,\Sigma ',q_0,F,\delta')$ as below:  
\begin{itemize}
    \item $\Sigma '=\{(a,\alpha,r) \}$ where $a\in \Sigma $, $\alpha \in \{p(h_1),!p(h_1) \}$, 
    $r \in \{\ins(h_1),- \}$. Thus $|\Sigma'|=4\times |\Sigma|$.
    The alphabet $\Sigma '$ contains all transitions that $M$ can have.
    \item $\delta'=\{(p,(a,\alpha,r),q) \:|\: (p,a,\alpha,r,q)\in \delta\}$
\end{itemize}

\noindent
From the construction of $M_1$, we see that it accepts all those possible sequences of transitions that may take $M$ from its initial state to any of its final states.
We construct NFA $M_2=(Q'',\Sigma ',q_0,Q'',\delta'')$ as below:  
\begin{itemize}
    \item $Q''=\{q_0,q_1\}$
   
\end{itemize}
\noindent

The transitions in $\delta''$
are as follows:
\begin{itemize}
    \item $\delta''(q_0,(a,!p(h_1),-))=\{q_0\}$, $a\in \Sigma $
    \item $\delta''(q_0,(a,!p(h_1),\ins(h_1)))=\{q_1\}$, $a\in \Sigma $
    \item $\delta''(q_1,(a,!p(h_1),\ins(h_1)))=\{q_1\}$, $a\in \Sigma $
    \item $\delta''(q_1,(a,p(h_1),\ins(h_1)))=\{q_1\}$, $a\in \Sigma $
    \item $\delta''(q_1,(a,p(h_1),-))=\{q_1\}$, $a\in \Sigma $
    \item $\delta''(q_1,(a,!p(h_1),-))={q_1}$, $a\in \Sigma $
\end{itemize}

\noindent 
The automaton $M_2$ works as follows. 
State $q_0$ denotes that we have not yet encountered $\ins(h_1)$, and state $q_1$ denotes we have seen an $\ins(h_1)$.  
 For inputs of the form $(x,!p(h_1),-)$ at $q_0$ where $x\in \Sigma $, we remain in state $q_0$. If we come across inputs of the form $(x,!p(h_1),\ins(h_1))$ where $x\in \Sigma $, we move to state $q_1$ from $q_0$.  At state $q_1$, the automaton remains in state $q_1$ for every element in $\Sigma '$.

By construction, $M_2$ accepts all those sequences of transitions of $M$ where every transition containing $p(h_1)$ is preceded by at least one transition containing $\ins(h_1)$.
The NFA $M_3$ is a synchronous product of NFA $M_1$ and NFA $M_2$. Therefore, $M_3$ accepts all those sequences which take the 
SAFA $M$ from its initial state to a final state. 
Thus, if the language accepted by $M_3$ is empty, so is the language accepted by the SAFA $M$, and is non-empty 
otherwise. 
The automaton $M_3$ is non-empty if a simple path exists from its initial state to any of its final states. The simple path can be found using a standard Depth First Search (DFS).

The time complexity of the emptiness check is polynomial in the size of $M_3$. 
The size of $M_3$ depends on the size of $M_1$ and $M_2$.
The number of states in $M_1=|Q|$ is the same as $M$. The number of states in $M_2$ is $2$, the number of states in $M_3$ is at most $2|Q|$, and the number of edges in $M_3$ is at most $O(|\Sigma |\times |\delta|\times |Q|)$ which is polynomial in the input size.
\end{proof}
\begin{example}\label{egdec}
 The NFA $M_1$, $M_2$, $M_3$  corresponding to SAFA $M$ (Figure \ref{f4}) are shown in Figures \ref{f5}, \ref{f6}, and \ref{f7} respectively. 
 We observe that in NFA $M_3$, there exists a path from the initial to the final state. Hence, NFA $M_3$ is not empty, and therefore the SAFA $M$ in Figure \ref{f4} is also not empty, which is indeed the case.
\qed
\end{example}

\begin{theorem}\label{hisnlcomp}
   The nonemptiness problem for singleton SAFA is $\NL$-complete.
\end{theorem}
\begin{proof}
We first discuss $\NL$-membership. 
By Lemma~\ref{lem:SAFAtoNFA}, the nonemptiness for singleton SAFA $M=(Q,\Sigma \times D,q_0,F,H,\delta)$ is in ${\sf PTIME}$ by reducing the problem to checking the nonemptiness of a nondeterministic finite automaton (NFA) with $2|Q|$ states. 
This NFA can be constructed on-the-fly leading to an $\NL$-membership of the nonemptiness problem of singleton SAFA.

For $\NL$-hardness, we show a reduction from the reachability problem 
 on a directed graph $G$ having a vertex set $V=\{1,\dots n\}$ which is known to be $\NL$-complete~\cite{jones1975space}.
Let $G$ be a directed graph with $V=\{1,2,...,n\}$, and we are given the vertices $1$ and $n$ as source and target vertices, respectively.
We define a SAFA $M=(Q,\Sigma\times D,q_0,F,H,\delta)$ where $Q=V$, $\Sigma=\{a\}$, $D$ is a countably infinite set, $q_0=1$, $F=\{n\}$, $H=\{h_1\}$ i.e. $|H|=1$. The transitions in $\delta$ are as follows:
$(i,a,!p(h_1),-,j) \in \delta$ if $(i,j)$ is an edge in $G$.
It is easy to see that $M$ can be constructed from $G$ in logspace
and that $L(M)\neq \phi$ if and only if there is a path from vertex $1$ to vertex $n$ in $G$. Hence, the result.
\end{proof}

\noindent
We now show that membership for SAFA is \NP-complete.
We first note that, unlike the nonemptiness problem, for DSAFA membership can be decided in $\PTIME$ by reading the input word and checking if a final state is reached.

\begin{theorem}\label {member}
The membership problem for SAFA is \NP-complete.
\end{theorem}
\begin{proof}
Given a SAFA $M$ and an input word $w$, if $w \in L(M)$, then a nondeterministic Turing machine can guess an accepting run in polynomial time, and hence, the membership problem is in \NP.

For showing \NP-hardness, we reduce from $3$SAT for the nonemptiness problem as done for Lemma~\ref{safa_hard}.
Instead of a deterministic automaton that was constructed in the proof of Lemma~\ref{safa_hard}, we construct a nondeterministic SAFA $M$ with $\Sigma=\{a\}$, and all the transitions in $M$ are labelled with the same letter $a \in \Sigma$.
Everything else remains the same as in the construction in Lemma~\ref{safa_hard}.
Note that in the 3SAT formula $\psi$, if there are $\ell$ variables and $k$ clauses, then there is a path of length $\ell + k$ from the initial state to the unique final state of $M$.
We consider an input word $w=(a,d) \cdots(a,d)$, that is a word in which all the attribute, data-value pairs are identical in the whole word such that $|w| = \ell + k$.
It is not difficult to see that $w \in L(M)$ if and only if $\psi$ is satisfiable.
\end{proof}

Finally, we show that given a SAFA $M$ defined on $\Sigma \times D$, whether $L(M)=(\Sigma \times D)^*$ (universality problem) is undecidable. We show that the universality problem is undecidable even for singleton SAFA. This is in contrast with the universality problem for register automata where one needs at least two registers to prove undecidabilty of the universality problem~\cite{6}. Note that the universality problem is decidable for register automata with one register~\cite{10}. However, the problem is undecidable for guessing register automata even with one register~\cite{czerwinski2021new}.  

\begin{theorem}\label{thm:univ}
The universality problem for singleton SAFA is undecidable.
\end{theorem}

\begin{proof}
This proof is line with the proof in \cite{czerwinski2021new,bojanczyk2019slightly} for universality of singleton guess register automata. We reduce the halting problem for two-counter machines (also known as Minsky machines for which the halting problem is already known to be undecidable) to the non-universality problem for singleton SAFA. Let $\mathcal{M} = (Q,\delta,q_0,q_f)$ be a two-counter machine with counters $C_1$ and $C_2$ both of which are initially set to $0$, a set $Q$ of states, a set $\delta$ of transitions, $q_0$ is the initial state, and $q_f$ is the halting state.

\begin{enumerate}
    \item Transitions which increment a counter and are of the form $(q_i,inc(C_j),q_k)$ that take the machine from state $q_i$ to $q_k$, and increase the value of counter $C_j$ by $1$, where $q_i,q_k\in Q$ and $C_j\in \{C_1,C_2\}$.
    \item Transitions which decrement a counter and are of the form $(q_i,dec(C_j),q_k)$ that take the machine from state $q_i$ to $q_k$, and decrease the value of counter $C_j$ by $1$, where $q_i,q_k\in Q$ and $C_j\in \{C_1,C_2\}$.
    \item Transitions which jump on $0$ and are of the form $(q_i,jz(C_j),q_k)$ that take the machine from state $q_i$ to $q_k$ only if the value of the counter $C_j$ is $0$ where $q_i,q_k\in Q$ and $C_j\in \{C_1,C_2\}$.
\end{enumerate} 
Given a state, either it has an increment transition or a decrement transition associated with it. The decrement transition states can also have jump on $0$ transition. A Minsky machine halts iff the sequence of transitions ends in a halting state and value of both its counters is $0$.
We construct a nondeterministic singleton SAFA $M$ such that $M$ accepts a data word that represents a sequence of transitions of a two-counter machine $\mathcal{M}$ if and only if it is not in the correct format or is not a valid halting computation of $\mathcal{M}$. Consequently, $\mathcal{M}$ halts if and only if $L(M)$ is non-universal.

\subsection*{1. Encoding Computations as Data Words}
A data word represents a sequence of transitions of the Minsky machine. Suppose that the Minsky machine begins in state $q_0$ with counter values $0$ for both $C_1$ and $C_2$ respectively. Suppose, a transition $(q_0, inc(C_1), q_2)$ is applied to the initial configuration. Suppose the transition $(q_2, dec(C_1), q_3)$ is applied next. Let the last transition applied be $(q_3, jz(C_1), q_f)$. Then the corresponding transition sequence is $(q_0, inc(C_1), q_2)\\(q_2, dec(C_1), q_3)(q_3, jz(C_1), q_f)$ where $q_f$ is the halting state of the Minsky Machine.

For the construction of the data word, we consider $\Sigma = \delta \cup \{\$\}$. Here the symbol $\$$ is both the beginning and the end marker. A possible data word corresponding to the sequence of transitions above is $(\$,109)((q_0, inc(C_1), q_2),47)\\((q_2, dec(C_1), q_3),47)((q_3, jz(C_1), q_f),89)(\$,67)$. 

The corresponding data word is constructed from the transition sequence using the following rules.
\begin{enumerate}
\item The data word begins and ends, with the end marker $(\$,d)$ where $d\in \mathbb{N}$.
\item All the increment, decrement and jump on zero transitions must be assigned data values. 
\end{enumerate}

A data word $w$ represents a valid halting transition sequence in the required format if it satisfies the following properties:
\begin{enumerate}
    \item The data word begins and ends with the end marker $(\$,d)$ where $d\in \mathbb{N}$.
    \item The first transition has $q_0$ as source state and the last transition in the sequence of transitions has $q_f$ as the target state. This ensures the Minsky machine started in the initial state and ended in a halting state.
    \item For all transitions, the target state of a transition matches the source state of the next transition in the sequence. This ensures the sequence of transitions is valid.
    \item No two increment transitions are assigned the same data value. As we will see in point~\ref{itm:inc_dec}, in a valid halting transition sequence, for every increment operation there needs to be a corresponding decrement operation. The increment-decrement pair is identified by data values. If more than one increment operation had the same data value then all of them could correspond to a single decrement operation having the same data value, which would lead to simulating an invalid sequence of transitions.\label{itm:uniq_inc}
    \item No two jump on zero transitions are assigned the same data value. This enables us to identify each jump on zero instruction uniquely.
    \item All decrement transitions are assigned data values which have already been assigned to increment operations which came prior to them in the sequence of transitions. Moreover, no two decrement transitions are assigned the same data values. The data values associated with decrement transitions are unique due to the same reason as uniquneness of data values associated with increment transitions (see point \ref{itm:uniq_inc}).
    \item  Every increment operation on a counter, must have a corresponding decrement operation on the same counter with the same data value as the increment counter. Moreover, there cannot be any jump on zero transition on the same counter in between the increment and decrement operations. This ensures at the end of the transition sequence the value of both the counters are $0$ and also ensures jump on zero has not been executed erroneously. \label{itm:inc_dec}
    
\end{enumerate}

\subsection*{2. Construction of the singleton SAFA $M$}
The automaton $M$ is constructed to nondeterministically guess and verify a violation of at least one of the above properties.
\begin{itemize}
\item A sequence of transitions either not beginning or not ending in $(\$,d)$ where $d\in \mathbb{N}$ can be accepted by an NFA. 
\item A sequence of transitions with the first transition not having $q_0$ as source state or the last transition not having $q_f$ as the target state can again be accepted by an NFA.
\item Two adjacent transitions whose source and target state do not match can be identified once again by an NFA.
\item A sequence of transitions with more than one increment transition being assigned the same data value can be identified by using a singleton SAFA which stores the data values of the increment transitions in a set, and if a data value is repeated, the SAFA accepts the data word.
\item A sequence of transitions with more than one jump on zero transition being assigned the same data value can be identified by using a singleton SAFA which stores the data values of the jump on zero transitions in a set, and if a data value is repeated, the SAFA accepts the data word.
\item A sequence of transitions with a decrement transition being assigned a data value that is not assigned to any increment operation coming prior to it can again be accepted by a singleton SAFA. The SAFA stores the data values associated with the increment transitions in the data word in a set. When it comes across a decrement transition whose corresponding data value is not present in the set, it accepts the data word.
\item More than one decrement transition being assigned the same data value can be identified by using a singleton SAFA which stores the data values of the decrement transitions in a set, and if a data value is repeated, the SAFA accepts the data word.
\item If an increment transition does not have a corresponding decrement transition following it having the same data value, the data word is accepted or if their is a jump on zero on the same counter as the increment operation before a decrement operation on the same counter, then the data word is also accepted. 
These again can be identified using a singleton SAFA, which nondeterministically guesses the increment transition which does not have a corresponding decrement transition, or prior to getting the decrement transition, there is a jump on zero in between and stores that data value associated with the increment operation in the set. 
If after scanning the whole data word, the singleton SAFA does not find a decrement transition with the same data value or if it finds a jump on zero on the same counter as the increment operation before it comes across the corresponding decrement transition, it accepts the data word.
\end{itemize}
\noindent
All the above checks can either be done by an NFA or by a singleton SAFA.
The SAFA $M$ is the union of all the above NFAs and singleton SAFAs. Since singleton SAFA are closed under union, we have that $M$ is singleton. The singleton SAFA $M$ nondeterministically guesses the violation in the data word and then checks for it.

\subsection*{3.Reduction}
Let us analyze $L(M)$. 
\begin{itemize}
    \item \textbf{Case 1: $\mathcal{M}$ halts.} There exists at least one valid sequence of transitions $w_{valid}$ in the required format (Recall that although the two-counter machine is deterministic and there is only one halting computation if it halts, but when the computation is converted to a data word, various permutations of data values can be assigned. Hence there can be more than one data word which represents the halting computation). Because $w_{valid}$ violates none of the properties, every non-deterministic branch of the singleton SAFA $M$ fails to detect an error and thus rejects $w_{valid}$. Thus, $w_{valid} \notin L(M)$, meaning $L(M) \subsetneq (\Sigma \times \mathbb{N})^*$. The language of M is thus not universal. 
    \item \textbf{Case 2: $\mathcal{M}$ does not halt.} No valid halting sequence of transition exists. Therefore, every possible data word $w \in (\Sigma \times \mathbb{N})^*$ contains at least one violation of the properties. Since $M$ detects all possible violations, at least one non-deterministic branch of $M$ will accept $w$. Thus, we have that $L(M) = (\Sigma \times \mathbb{N})^*$ and hence universal. 
\end{itemize}

We have shown that $\mathcal{M}$ halts if and only if $L(M)$ is non-universal. Since the halting problem for two-counter machines is undecidable, it follows that the universality problem for singleton SAFA is undecidable.
\end{proof}

\subsection{Closure Properties}\label{close}
We now study the closure properties.
We first study the closure properties of SAFA followed by those of DSAFA.

\subsubsection*{Closure Properties of SAFA}
We start with the Boolean closure properties.
We show that SAFA are closed under union, but not under intersection and complementation.

\begin{lemma}\label{lem:union}
SAFA are closed under union.
\end{lemma}

\begin{proof}

We show that SAFA are closed under union.
The union of two SAFAs can be obtained by superimposing their start states. Let us consider two SAFA  $M_1 = (Q_1,\Sigma \times D, q_{01}, F_1, H_1, \delta_1)$ and  $M_2 = (Q_2,\Sigma \times D, q_{02}, F_2, H_2, \delta_2)$. The SAFA $M_3=(Q_3,\Sigma \times D, q_{03}, F_3, H_3, \delta_3)$ which accepts the language $L(M_1)\cup L(M_2)$ is constructed as follows:
$Q_3=\{q_{03}\}\cup Q_1 \cup Q_2$, $F_3=F_1 \cup F_2$ if $q_{01}\notin F_1$ and $q_{02}\notin F_2$, otherwise $F_3=F_1 \cup F_2 \cup \{q_{03}\}$, $H_3=H_1 \cup H_2$.
All transitions in $\delta_1$ and $\delta_2$ are in $\delta_3$. Additionally, for every transition $(a,\gamma,\op)$ from state $q_{01}$ to any state $q_i\in Q_1$ in $\delta_1$ such that $a\in \Sigma $, $\gamma \in \{p(h_k),!p(h_k)\}$ with $h_k\in H_1$ and $\op \in \{-, \ins(h_l)\}$, $h_l\in H_1$, the transition $(a,\gamma,\op)$ is included in $\delta_3$ from state $q_{03}$ to state $q_i$. Similarly, for every transition $(a,\gamma,\op)$ from state $q_{02}$ to a state $q_i\in Q_2$ in $\delta_2$ such that $x\in \Sigma $, $\gamma\in \{p(h_k),!p(h_k)\}$,  $h_k\in H_2$ and $\op \in \{-, \ins(h_l)\}$, $h_l\in H_2$,  the transition $(a,\gamma,\op)$ is included in $\delta_3$ from state $q_{03}$ to state $q_i$. 
The automaton $M_3$ on an input data word nondeterministically decides to which of $M_1$ and $M_2$ the input word belongs. 
If either automaton accepts the input word, it is accepted by $M_3$.
If both the automata do not accept an input word, it is rejected. Thus, the SAFA $M_3$ accepts the language $L(M_1) \cup L(M_2)$. 
\end{proof}

\begin{lemma} \label{lem:intersection}
SAFA are not closed under intersection.    
\end{lemma}

\begin{proof}
Consider the language $L=L_{{\sf fd}(a)}\cap L_{a\exists b}$. We show that there exists no SAFA which accepts $L$. We prove by contradiction. Assume that there exists a SAFA $M=(Q,\{a,b\} \times D,q_0,F,H,\delta)$ with $|H|=k>0$ such that $M$ accepts $L$. 
Then $M$ must accept the following word $w \in L$ where
$w=(b,d_1)\cdots (b,d_{k+1})(a,d_1)\cdots (a,d_{k+1})$ and $d_1,...,d_{k+1}$ are all distinct.
In order to accept $w$, the SAFA $M$ must go through a sequence $T=t_{b_1}...t_{b_{k+1}}t_{a_1}...t_{a_{k+1}}$ of transitions  to completely consume $w$ and end in an accepting state.
Here, $t_{b_i}$ consumes the data element $(b,d_i)$, and $t_{a_j}$ consumes the data element $(a,d_j)$ and $1\leq i,j\leq k+1$. 
Two cases are possible:
\begin{itemize}
    \item There is a transition say $t_{a_g}$ in $T$ consuming the data element $(a,d_g)$ of $w$ where $g\in [k+1]$ and $t_{a_g}$ is of the form $(a,!p(h_i),-)$ or $(a,!p(h_i),\ins(h_j))$ where $h_i,h_j\in H$. 
    The SAFA $M$ using the same sequence $T$ of transitions can accept another data word $w' \notin L_{a\exists b}$ where $(a,d_g)$ is replaced by $(a,d)$ such that $d \neq d_r$ for $1 \leq r \leq k+1$. It is always possible to get such a data value $d$ as $k$ is finite and $D$ is countably infinite. 
    . Therefore, at the time of executing $t_{a_g}$ the data value $d$ is not present in $h_i$, and $t_{a_g}$ executes successfully. Recall that the data values in $w$ that follow $d_j$ are different from $d_j$. Also, by the choice of $d$, the data values in $w'$ that follow the data value $d$ are not equal to $d$.
    Therefore, whether $d$ has been inserted into any set $h_j\in H$ or not while executing $t_{a_g}$ does not affect the successful execution of the transitions in $T$ that follow $t_{a_g}$. Now, in $w'$, there is a data value $d$ associated with attribute $a$, which is not associated with attribute $b$, thus $w'\notin L_{a \exists b}$.

    \item All transitions in $T$ following $t_{b_{k+1}}$ are of the form $(a,p(h_i),-)$ or $(a,p(h_i),\ins(h_j))$ where $h_i,h_j\in H$. 
    The number of transitions in $T$ that follow $t_{b_{k+1}}$ is greater than $k$.
    Hence, by the pigeonhole principle, there must be two transitions $t_{a_\ell}$ and $t_{a_m}$ that use the same set $h_i$ in the condition $p(h_i)$ for some $h_i \in H$, and $1\leq \ell < m\leq k+1$. 
    The SAFA $M$, using the same sequence $T$ of transitions, can accept another data word $w' \notin L_{{\sf fd}(a)}$ in which $(a,d_m)$ is replaced by $(a,d_l)$. 
    The SAFA $M$, when executing $t_{a_m}$ on $w'$, can successfully consume the data element $(a,d_\ell)$ instead of $(a,d_m)$.
    This is because $t_{a_\ell}$ and $t_{a_m}$ have the same condition $p(h_i)$ and $d_\ell$ is already present in $h_i$ when $t_{a_\ell}$ is executed. 
    Note that the data values in $w'$ that follow the execution of $t_{a_m}$ are not equal to either $d_\ell$ or $d_m$.
    Therefore, whether $d_\ell$ has been inserted into any set $h_j\in H$ or not does not affect the successful execution of the transitions in $T$ that follow $t_{a_m}$.
\end{itemize}
Thus, $M$ accepts both $w\in L$ and $w'\notin L$, hence the contradiction.
\end{proof}

Lemma~\ref{lem:complementation} follows from Lemma~\ref{lem:intersection} and~\ref{lem:union}.
\begin{lemma} \label{lem:complementation}
SAFA are not closed under complementation.    
\end{lemma}

\begin{theorem}
SAFA are closed under concatenation.
\end{theorem}

\begin{proof}
SAFA are shown to be closed under concatenation in a manner similar to NFA.
Note that if the sets used in the two input automata $A$ and $B$ be $H_A$ and $H_B$, respectively, then the automaton accepting the language obtained as a result of concatenation uses the set $H_A \cup H_B$.
\end{proof}

\begin{theorem} \label{thm:kleene}
SAFA are not closed under Kleene's closure.
\end{theorem}

\begin{proof}
Consider the language 
$L_1=\{(a,d)(a,d) \:|\: d\in D\}$. The language $L_1$ can be accepted by a DSAFA (see Figure~\ref{fig_kleene}). The Kleene's closure of $L_1$ is the language $L={L_1}^*$,
 that is, $L$ is the set of all data words where every data value appears in pairs. We show that there exists no SAFA which accepts $L$. We prove by contradiction. Assume that there exists a SAFA $M=(Q,\{a\} \times D,q_0,F,H,\delta)$ with $|H|=k>0$ such that $M$ accepts $L$. 
Then $M$ must accept the following word $w \in L$ where
$w=(a,d_1)(a,d_1)\cdots(a,d_i)(a,d_i)\cdots (a,d_{k+1})(a,d_{k+1})$ and $d_1,...,d_{k+1}\in D$ are all distinct.
 
In order to accept $w$, the SAFA $M$ must go through a sequence $T=t_{1_{d_1}}t_{2_{d_1}}...t_{1_{d_{k+1}}}t_{2_{d_{k+1}}}$ of transitions to completely consume $w$ and end in an accepting state.
Here, $t_{1_{d_i}}$ consumes the first data item of the $i^{th}$ data value pair, and $t_{2_{d_i}}$ consumes the second data item of the $i^{th}$ data value pair and $1\leq i\leq k+1$. We note the following:

\begin{itemize}
   \item The transitions $t_{2_{d_i}}$ must be of the form $(a,p(h_\ell),-)$ or $(a,p(h_\ell),\ins(h_j))$ where $h_\ell,h_j\in H$. This is essential because if $t_{2_{d_i}}$ is of the form $(a,!p(h_\ell),-)$ or $(a,!p(h_\ell),\ins(h_j))$ then instead of consuming $d_i$ it can also consume successfully a new data value $d_{new}\neq d_i$. It is always possible to get such a data value as $D$ is countably infinite. The SAFA $M$ will then accept the data word $w'=(a,d_1)(a,d_1)\cdots(a,d_i)(a,d_{new})\cdots \\(a,d_{k+1})(a,d_{k+1})$ which is not in $L$.

   \item The transitions $t_{1_{d_i}}$ must be of the form $(a,!p(h_\ell),\ins(h_j))$ where $h_\ell,h_j\in H$. This is essential because if $t_{1_{d_i}}$ is of the form $(a,p(h_\ell),-)$ or $(a,p(h_\ell),\ins(h_j))$, then $t_{1_{d_i}}$ will fail to consume the first instance of the data value $d_i$. Recall that the data values $d_1,\dots ,d_{k+1}$ are all distinct. 
   Further, the transition $t_{1_{d_i}}$ cannot be of the form $(a,!p(h_\ell),-)$, otherwise the following transition $t_{2_{d_i}}$ which is of the form $(a,p(h_j),-)$ or $(a,p(h_j),\ins(h_p))$ where $h_j, h_p\in H$ will not be executed successfully. 
   Note that for the transition $t_{2_{d_i}}$ to consume the data value $d_i$, the data value $d_i$ must be inserted in a set when it was first encountered along $t_{1_{d_i}}$.

   \item Since the number of distinct data values in $w$ is more than the number of sets in $M$ and all the distinct data values are inserted in the sets in $M$, by the pigeonhole principle, there are two distinct data values $d_i$ and $d_j$, with $i<j$ which are inserted in the same set $h_\ell \in H$. Thus, if $M$ accepts the data word $w=(a,d_1)(a,d_1)\cdots(a,d_i)(a,d_i)\cdots \\(a,d_j)(a,d_j)\cdots (a,d_{k+1})(a,d_{k+1})$ then $M$ also accepts the data word 
   
   $w'=(a,d_1)(a,d_1)\cdots(a,d_i)(a,d_i)\cdots (a,d_j)(a,d_i)\cdots (a,d_{k+1})(a,d_{k+1})$ \\which is not in $L$.
\end{itemize}
   
    Thus, SAFA are not closed under Kleene's closure.
\end{proof}
\begin{figure} [t]
\begin{minipage}{0.55\textwidth} 
\centering
\begin{tikzpicture} [node distance = 3.6cm, on grid, auto]

\node (q0) [state, initial,  initial text = {}] {$q_0$};
\node (q1) [state, right = 3.8cm of q0] {$q_1$};
\node (q2) [state, accepting, right = 3cm of q1] {$q_2$};

\path [-stealth, thick]
	(q0) edge  node[above] {{\footnotesize $(a,!p(h_1),\ins(h_1))$}}    (q1)
	
    (q1) edge  node[above] {{\footnotesize $(a,p(h_1),-)$}}    (q2);
	
\end{tikzpicture}
\begin{centering}
\caption{A SAFA $M$, such that $L(M)^*\notin \lsafa$} \label{fig_kleene}
\end{centering}
\end{minipage}
\hfill
\begin{minipage}{0.35\textwidth} 
\centering
\begin{tikzpicture} [node distance = 3cm, on grid, auto]

\node (q0) [state, initial, accepting, initial text = {}] {$q_0$};
\node (q1) [state, right = of q0] {$q_1$};

\path [-stealth, thick]
	(q0) edge [loop below]  node [right] {{\footnotesize $(a,p(h_1),-)$}}    (q0)
	(q0) edge [loop above]  node [right] {{\footnotesize $(b,p(h_1),-), (b,!p(h_1),\ins (h_1)) $}}    (q0)

    (q0) edge  node[above] {{\footnotesize $(a,!p(h_1),-)$}}   (q1);
	
\end{tikzpicture}
\begin{centering}
\caption{DSAFA for $L_{\sf a\exists b}$} \label{fig_aeb}
\end{centering}
\end{minipage}
\end{figure}

\begin{theorem}
SAFA are not closed under reversal.
\end{theorem}
\begin{proof}
Consider the language $L_{a\exists b}$, which can be accepted by a DSAFA (see Figure~\ref{fig_aeb}). The reversal of $L_{a\exists b}$ is the language $L=\{w^R \:|\: w\in L_{a\exists b}\}$ i.e. $L$ is the set of all data words where for every attribute $a$ there is an attribute $b$ which comes after it and whose data value is same as that of attribute $a$. We show that there exists no SAFA which accepts $L$. We prove by contradiction. Assume that there exists a SAFA $M=(Q,\{a,b\} \times D,q_0,F,H,\delta)$ with $|H|=k>0$ such that $M$ accepts $L$. 
Then $M$ must accept the following word $w \in L$ where
$w=(a,d_1)\cdots (a,d_{k+1})(b,d_1)\cdots (b,d_{k+1})$ and $d_1,...,d_{k+1}\in D$ are all distinct. 
In order to accept $w$, the SAFA $M$ must go through a sequence $T=t_{a_{d_1}}...t_{a_{d_{k+1}}}t_{b_{d_1}}...t_{b_{d_{k+1}}}$ of transitions to completely consume $w$ and end in an accepting state.
Here, $t_{a_{d_i}}$ consumes the data item $(a,d_i)$ and $t_{b_{d_i}}$ consumes the data item $(b,d_i)$ of $w$. Now, we have the following observations.
\begin{itemize}
   \item The transitions $t_{b_{d_i}}$ must be of the form $(b,p(h_\ell),-)$ or $(b,p(h_\ell),\ins(h_j))$ where $h_\ell,h_j\in H$. This is essential because if $t_{b_{d_i}}$ is of the form $(b,!p(h_\ell),-)$ or $(b,!p(h_\ell),\ins(h_j))$ then instead of consuming $d_i$ it can also consume successfully a new data value $d_{new}\in D$ which is not present in $w$. It is always possible to get such a data value as $D$ is countably infinite. The SAFA $M$ will then accept the data word \\$w'=(a,d_1)\cdots (a,d_i)\cdots (a,d_{k+1})(b,d_1)\cdots (b,d_{new})\cdots (b,d_{k+1})$ which is not in $L$.

   \item The transitions $t_{a_{d_i}}$ must be of the form $(a,!p(h_\ell),\ins(h_j))$ where $h_\ell,h_j\in H$. This is essential because if $t_{a_{d_i}}$ is of the form $(a,p(h_\ell),-)$ or $(a,p(h_\ell),\ins(h_j))$, then $t_{a_{d_i}}$ will fail to consume the first instance of the data value $d_i$ since all data values in $w$ associated with attribute $a$ are distinct. Further, The transition $t_{a_{d_i}}$ cannot be of the form $(a,!p(h_\ell),-)$ because there is a following transition $t_{b_{d_i}}$, which is of the form $(b,p(h_j),-)$ or $(b,p(h_j),\ins(h_p))$ where $h_j, h_p\in H$ that will not be executed successfully. Recall that for the transition $t_{b_{d_i}}$ to consume the data value $d_i$, the data value $d_i$ must be inserted in a set when it was first encountered.

   \item Since the number of distinct data values in $w$ associated with attribute $a$ is more than the number of sets in $M$ and all the distinct data values associated with attribute $a$ are inserted in the sets in $M$, by the pigeonhole principle, there are two distinct data values $d_i$ and $d_j$, with $i<j$ which are inserted in the same set $h_\ell \in H$. Thus, if $M$ accepts the data word $w=(a,d_1)\cdots(a,d_i)\cdots (a,d_j)\cdots (a,d_{k+1})(b,d_1)\cdots(b,d_i)\cdots (b,d_j)\cdots \\(b,d_{k+1})$ then $M$ also accepts the data word 
   
   $w'=(a,d_1)\cdots(a,d_i)\cdots (a,d_j)\cdots (a,d_{k+1})(b,d_1)\cdots(b,d_i)\cdots (b,d_i)\cdots \\(b,d_{k+1})$ which is not in $L$.
\end{itemize}   
    Thus, SAFA are not closed under reversal.   
\end{proof}
\begin{theorem}
 SAFA are not closed under homomorphism.
\end{theorem}
\begin{proof}
 Consider the language $L_1=\{\{a\}\times D\}^*$. The homomorphism function is $h(\epsilon)=\epsilon$, $h((a,d))=(a,d)(a,d)$ for all $d\in D$. 
The language $L=h(L_1)$ is the set of all data words where every data value appears in
pairs. It is already shown in Theorem~\ref{thm:kleene} that SAFA cannot accept $L$.
\end{proof}

\begin{theorem}
 SAFA are not closed under inverse homomorphism.
\end{theorem}
\begin{proof}
Consider the language $L_\epsilon=\{\epsilon \}$. There exists a SAFA $M$ which accepts $L_\epsilon$. The data values are taken from the set of natural numbers $\mathbb{N}$ and $\Sigma=\{a\}$. The homomorphism function is $h(\epsilon)=\epsilon$, $h((a,1))=\epsilon$, $h((a,d))=(a,d)$ for all $d\in \mathbb{N}\setminus \{1\}$. The language $L=h^{-1}(L_\epsilon)$ is a language of data words where the data word is $\epsilon$ or data words that only have data value $1$ present in it. 
We prove by contradiction. Assume that there exists a SAFA $M=(Q,\{a\} \times D,q_0,F,H,\delta)$ with $|H|=k>0$ such that $M$ accepts $L=h^{-1}(L_\epsilon)$.
Then $M$ must accept the following data word $w \in L$ where $w=(a,1)$.
 
In order to accept $w$, the SAFA $M$ must go through a transition $t$  to consume $w$ and end in an accepting state.

We note the following:

\begin{itemize}
   \item The transitions $t$ must be of the form $(a,!p(h_i),-)$ or $(a,!p(h_i),\ins(h_j))$ where $h_i,h_j\in H$. Since, $t$ is the first transition it cannot be of the form $(a,p(h_i),-)$ or $(a,p(h_i),\ins(h_j))$ because all the sets of SAFA $M$ are empty at the beginning and $p(h_i)$ will always fail. If $t$ is of the form $(a,!p(h_i),-)$ or $(a,!p(h_i),\ins(h_j))$ then it can also consume the data word $(a,2)$ and go to a final state as $2$ is also not present in $h_i$ as $t$ is the first transition and all the sets of SAFA $M$ are empty. Thus $M$ accepts the dataword $(a,2)$, but the data word $(a,2)\notin L=h^{-1}(L_\epsilon)$.
\end{itemize}
    Thus, SAFA are not closed under inverse homomorphism.
\end{proof}

\subsubsection*{Closure properties of Deterministic SAFA:}
Here, we discuss deterministic SAFA, their closure properties, and compare their expressiveness and succinctness with SAFA.
Using standard complementation construction as in deterministic finite automata (DFA), by changing non-accepting states to accepting and vice versa, we can show that DSAFA are closed under complementation. Moreover, as deterministic SAFA are closed under complementation but not under intersection (see Lemma~\ref{lem:intersection}), it follows that they are also not closed under union. 
Since the languages used to show non-closure of SAFA under Kleene's closure, homomorphism, and inverse homomorphism are accepted by DSAFA, we have that DSAFA are also not closed under Kleene's closure, homomorphism, and inverse homomorphism. 

\begin{theorem}
DSAFA are not closed under concatenation.
\end{theorem}

\begin{proof}
Consider the language $L_1=L_{{\sf fd}(a)}$ with  $\Sigma= \{a\}$ and $L_2=\{(a,d)(a,d)\\ \:|\: d\in D\}$. DSAFA can accept both $L_1$ and $L_2$. The concatenation of $L_1$ and $L_2$ is the language $L=L_1 \cdot L_2$. 
That is, $L$ is the set of all data words that end with a pair of the same data values, and all other data values present in the data word other than the data value in the pair are distinct. 
The frequency of the data value in the pair is either two or three. We show that there exists no DSAFA which accepts $L$.
When we see the first $(a,d)$ in the pair, a deterministic SAFA cannot determine if it is in $L_1$ or in $L_2$.

We prove by contradiction. Assume that there exists a DSAFA $M=(Q,\{a\} \times D,q_0,F,H,\delta)$ with $|Q|=n>0$ such that $M$ accepts $L$. 
Then $M$ must accept the following word $w \in L$ where
$w=(a,d_1)\cdots(a,d_n)(a,d_{n+1})\\(a,d_{n+1})$ and $d_1,...,d_{n+1}\in D$ are all distinct.
In order to accept $w$, the DSAFA $M$ must go through a sequence $T=t_1...t_{n+2}$ of transitions to consume $w$ and end in an accepting state. 
Let the run of $M$ on $w$ be a sequence $S=q_0...q_{n+2}$ of states to accept $w$ where $q_0,...,q_{n+2}\in Q$,  $q_{n+2}\in F$ and $q_0$ is the initial state. 
Here $t_i$ consumes $(a, d_i)$ and takes the DSAFA $M$ from state $q_{i-1}$ to $q_i$.

Since $|w|= n+2 > n=|Q|$, by the pigeonhole principle, some state $q_i$ for $i \in \{0, \dots, n+2\}$ in the sequence $S$ must be repeated in order to accept $w$.
We obtain the contradiction by showing that if any of the states in $S$ is repeated, then the DSAFA $M$ accepts a word $w'$ which is not in $L$.

We first show that the last state $q_{n+2}$ does not appear in the sequence earlier, i.e., for all $i \in \{0, \dots, n+1\}$, we have that $q_i \neq q_{n+2}$.
Otherwise, since $q_{n+2}$ is accepting, the DSAFA $M$ accepts a data word $w'\in L_{{\sf fd}(a)}$ which, is not in $L$. We argue that the state $q_{n+1}$ also does not appear earlier in the sequence $S$.
Note that since the data values $d_1, \dots, d_{n+1}$ are all distinct, the transitions $t_1,...,t_{n+1}$ must be of the form $(a,!p(h_\ell),-)$ or $(a,!p(h_\ell),\ins(h_j))$ where $h_\ell,h_j\in H$.
Now, if a state $q_i$ for $i < n+1$ is the same as $q_{n+1}$, then $M$ can execute the sequence $T_{new}=t_1...t_it_{i+1}...t_{n+1}t_{i+1}...t_{n+1}t_{n+2}$ of transitions and accept the data word $w'=(a,d_1)\cdots(a,d_n)(a,d_{n+1})(a,d_{new_{i+1}})\cdots$ $(a,d_{new_n})(a,d_{new_{n+1}})(a,d_{n+1})$ where $d_{new_{i+1}},...,d_{new_{n+1}}\in D$ are all distinct and also different from data values $d_1,...,d_{n+1}$.
The word $w' \notin L$ is accepted since the state reached after reading $w'$ is $q_{n+2} \in F$.
Hence, the state $q_{n+1}$ is not repeated in the sequence $S$.

Finally, we argue that none of the states in the sequence $q_0, \dots, q_{n}$ is also repeated.
Recall that the SAFA $M$ is a DSAFA, and in a DSAFA, from a state $q$, when $(a, d)$ is read for some $d$, there can be only two possible outgoing transitions from $q$; one with $p(h_\ell)$ and the other with $!p(h_\ell)$ for some $h_\ell \in H$.
If for some $i < j \leq n$, the state $q_j$ is such that $q_i = q_j$, then both the above outgoing transitions appear in the sequence $T$, one that follows the loop while the other one does not follow the loop and takes the run towards $q_{n+2}$.
Since this causes the transition with $p(h_\ell)$ to be taken from $q_i$ or $q_j$ in the run, the word $w$ has a data value $d \in \{d_1, \dots d_n\}$, which is repeated, thus contradicting the assumption that $d_1, \dots, d_n$ are all distinct in $w$.
Note that for $j=n$ in particular, the transition $t_{n+1}$ taken while reading $(a, d_{n+1})$ must have $!p(h_\ell)$ since $d_{n+1}$ is different from $d_k$ for $k \in [n]$.

   Hence, DSAFA are not closed under concatenation.   
\end{proof}

Thus, we get the following theorem. 
\begin{theorem}\label{dsafa_closure}
DSAFA are closed under complementation but not under union, intersection, concatenation, Kleene's closure, reversal, homomorphism and inverse homomorphism.
\end{theorem}
Table~\ref{closuretab} provides a summary of the closure properties of SAFA and DSAFA. We use the symbols $\cup$ for union, $\cap$ for intersection, $\overline{\phantom{x}}$ for complement, $\cdot$ for concatenation, $*$ for Kleene star, $L^R$ for reversal, $h(L)$ for homomorphism and $h^{-1}(L)$ for inverse homomorphism respectively.
\begin{table}[t]
\centering
\caption{Closure properties of SAFA}
\begin{tabular}{ |c|c|c|c|c|c|c|c|c| } 
 \hline
       & $\cup$ & $\cap$ & $\overline{\phantom{x}}$ & $.$ & $*$ & $L^R$ & $h(L)$ & $h^{-1}(L)$\\ 
       \hline
 SAFA & $\checkmark$ & $\times$ & $\times$ & $\checkmark$ & $\times$ & $\times$ & $\times$ & $\times$\\ 
 \hline
 DSAFA & $\times$ & $\times$ & $\checkmark$ & $\times$ & $\times$ & $\times$ & $\times$ & $\times$\\ 
  \hline

\end{tabular}
\label{closuretab}
\end{table}
Since SAFA is closed under concatenation but DSAFA is not, it follows that  the languages accepted by DSAFA is strictly contained in the class of languages accepted by SAFA.
\begin{theorem}\label{thm:dsafavsnsafa}
 $\ldsafa \subsetneq \lsafa$. 
\end{theorem}

Now, we discuss the succinctness between SAFA and DSAFA. Recall that $L_{\projs (L)=\regexp(r)}$ is the set of all data words $w$ such that $\projs(w)\in L_{\regexp(r)}$ where $L_{\regexp(r)} \subseteq \Sigma^*$ is the set of all words over $\Sigma$ generated by the regular expression $r$.
\begin{lemma}\label{dsafa=dfa}
    For every DSAFA accepting $L=L_{\projs (L)=\regexp(r)}$ we can always get a DFA accepting $L(r)$ where $L(r)$ is the language expressed by the regular expression $r$ which has the same number of states as the DSAFA.
\end{lemma}

\begin{proof}
 Consider the language $L=L_{\projs (L)=\regexp(r)}$. Let $M'$ be a DSAFA which accepts $L$. We can always obtain a DSAFA $M=(Q,\Sigma \times D,q_0,F,H,\delta)$ with a less or equal number of states as $M'$ and  transitions only of the form $(a,!p(h_i),-)$ where $a\in \Sigma$ and $ h_i,h_j\in H$, which accepts $L$ in the following manner:
     \begin{itemize}
         \item We can remove transitions of the form $(a,p(h_i),-)$ or $(a,p(h_i),\ins (h_j))$ where $a\in \Sigma$ and $ h_i,h_j\in H$ from $M'$ to construct $M$. 
         \item If there are transitions of the form $(a,!p(h_i),\ins (h_j))$ where $a\in \Sigma$ and $ h_i,h_j\in H$ in $M'$, we convert it to  $(a,!p(h_i),-)$ in $M$.
         \end{itemize}
         Thus, all transitions in $M$ are of the form $(a,!p(h_i),-)$.
         Observe that $M$ remains a DSAFA. The removal of $(a,p(h_i),-)$ or $(a,p(h_i),\ins (h_j))$ from $M'$ may result in some unreachable states in $M$ which can be removed. 
         $M$ accepts $L$ due to the following reasons:
         
         Consider the case that $M'$ has a sequence $T$ of transitions where there is at least one transition $t$ of the form $(a,p(h_i),-)$ which results in accepting a data word $w$ in $L$. The data word $w$ has $\projs(w)\in L(r)$ where $L(r)$ is the regular language expressed by the regular expression $r$ and $\projd(w)$ has at least one data value which is repeated. However, as $M'$ accepts $L$, it should also accept the data word $w'$ in $L$ where $\projs(w')=\projs(w)$ and $\projd(w')$ has data values which are all distinct. The sequence $T'$ of transitions which accepts $w'$ consists of transitions with $!p(h_i)$ where $h_i\in H$ only. The sequence $T'$ of transitions is also valid for $M$ as $M$ retains all transitions with $!p(h_i)$ and removes any insert operation if there are any from $M'$. This sequence $T'$ of transitions accepts both $w$ and $w'$ in $M$. 
     Given such a DSAFA $M$, we can get a DFA $A$, which accepts $L(r)$ by converting the transitions labelled $(a,p(h_i),-)$ to $a$ in $A$. The number of states in the DFA $A$ is the same as that of the DSAFA $M$.
\end{proof}

\begin{lemma}\label{dsafagdfa}
    Every DSAFA accepting $L=L_{\projs (L)=\regexp(r)}$ has at least  as many states as the smallest DFA accepting $L(r)$ where $L(r)$ is the language expressed by the regular expression $r$.
\end{lemma}

\begin{proof}
Let us assume that there exists a DSAFA $M$ which accepts $L=L_{\projs (L)=\regexp(r)}$ with $k_1$ states and there is a canonical minimal DFA~\cite{hopcroft2001introduction} $A_{min}$ with $k_2$ states which accepts $L(r)$ where $L(r)$ is the language expressed by regular expression $r$ such that $k_1<k_2$. 

From Lemma~\ref{dsafa=dfa} we see that there exists a DFA $A$ which accepts $L(r)$ with $k_1$ number of states. Thus, $A_{min}$ is no longer the minimal DFA, which is a contradiction.
\end{proof}

\begin{theorem}
There exists a language $L$ which is accepted by a nondeterministic SAFA with $n$ states but a DSAFA will require at least $2^{\Omega(n)}$ states to accept the same language $L$.
\end{theorem}

\begin{proof}
    From Lemma~\ref{dsafagdfa}, we see that the number of states of any DSAFA that accepts $L=L_{\projs (L)=\regexp(r)}$ must be greater than or equal to the number of states of the minimum DFA that accepts $L(r)$. Now consider a minimum NFA $A_{\mathit{nfa}}$, which accepts $L(r)$. From an NFA we can obtain a nondeterministic SAFA $M_{nfa}$, which accepts the language $L=L_{\projs (L)=\regexp(r)}$ by replacing the transitions of the NFA, which are of the form $a$ where $a\in \Sigma$ by the transition $(a,!p(h_i),-)$. The number of states of the nondeterministic SAFA $M$ is the same as that of the NFA $A_{\mathit{nfa}}$. We know from~\cite{hopcroft2001introduction} that there exists a family $\{L_n\}_{n\in \mathbb{N}}$ of languages such that if an NFA accepts $L_n$ with $n+1$ states then, every DFA accepting $L_n$ has size at least $2^n$. Let $r_n$ be the regular expression corresponding to language $L_n$. Using Lemma \ref{dsafagdfa}, the number of states of every DSAFA accepting $L=L_{\projs (L)=\regexp(r_n)}$ is at least $2^n$ while there exists a nondeterministic SAFA of $n+1$ states accepting $L=L_{\projs (L)=\regexp(r_n)}$.
\end{proof}

\section{Expressiveness}\label{sec:exp}

Let $\lkrfa$ and $\lcma$ be the set of all languages accepted by $k$-register automata and CMA, respectively.
We compare the computational power of SAFA with $k$-register automata. We show that $\lsafa$ and $\lkrfa$ are incomparable. Although $\lsafa$ and $\lkrfa$ are incomparable, SAFA recognizes many important languages which $k$-register automata also recognize,
such as $L_{d_1}$: wherein the first data value is repeated, $L_{a\geq 2}$: wherein attribute $a$ is associated with at least two distinct data values. On the other hand, $k$-register automata fail to accept languages where we have to store more than $k$ data values, such as $L_{{\sf fd}(a)}$, $L_{{\sf even}(a)}$: wherein attribute $a$ is associated with an even number of distinct data values~\cite{14}. 
SAFA can accept both these data languages. 
We also show below that there are languages such as $L_d$: the language of data words, each containing the data value $d$ associated with some attribute at some position in the data word, that can be accepted by a $2$-register (see Example~\ref{ra_pair}) but not by SAFA.

\begin{example}\label{ra_pair}
A $2$-register automaton can accept the language $L=\{(a,d)(a,d)\\|d\in D\}^*$.
Consider the $2$-register automaton $A$ with 
 $\Sigma =\{a\}$,
 $Q=\{q_0, q_1, q_2\}$,
 $\tau_0=\{\bot,\bot\}$,
 $F=\{q_0\}$,
 $U(q_0,a)=1$,
 $U(q_1,a)=2$, with initial state as $q_0$. The transition relation $\delta$ is defined as: $\{(q_0,a,1,q_1),(q_1,a,1,q_0), (q_1,a,2,q_2)\}$. The automaton $A$ accepts $L=\{(a,d)(a,d)|d\in D\}^*$. For an input word $w$, the automaton in state $q_0$ reads the first data value of a data pair and inserts it into register $1$ and goes to state $q_1$. 
In state $q_1$ if the second data value of the data pair is same as the first, the automaton goes back to state $q_0$ to read the next data value pair. 
Otherwise, the automaton inserts the data element in register $2$ and goes to a dead state $q_2$ and rejects the input word $w$. 
\qed
\end{example}

The following theorem follows from Example~\ref{ra_pair}, the fact that SAFA cannot accept the language $L=\{(a,d)(a,d) \:|\: d\in D\}^*$ (see theorem~\ref{thm:kleene}) and that $k$-register automata cannot accept $L_{\sf fd}(a)$~\cite{14}.

\begin{theorem}\label{t2}
$\lsafa$ and $\lkrfa$ are incomparable.
\end{theorem}

Remark: Since Bojanczyk's model~\cite{bojanczyk2020single} of single use register automata can also accept the language $L=\{(a,d)(a,d) \:|\: d\in D\}^*$ (the $2$-register automaton of Example~\ref{ra_pair} does not violate the single use register automaton conditions),  the set of languages accepted by SAFA and single use register automata~\cite{bojanczyk2020single} are also incomparable.

 We note that both SAFA and $k$-register automata have the same complexity for the nonemptiness and the membership problems.
Similar to SAFA, CCA and CMA also accept data languages such as $L_{\sf fd}(a)$ and $L_{{\sf even}(a)}$, which $k$-register automata cannot, but their decision problems have higher complexity~\cite{12,13}.
We can show that the class of languages accepted by SAFA is strictly contained in the class of languages accepted by CCA.
\begin{theorem}
$\lsafa \subsetneq \lcca$
\end{theorem}

\begin{proof}
For every SAFA $M=(Q,\Sigma \times D,q_0,F,H,\delta)$ with $|H|=k$ accepting a language $L$, we can construct a $k$-bag CCA $A=(Q,\Sigma,\Delta,\{q_0\},F)$, which accepts the same language $L$ in the following manner:
\begin{itemize}
    \item The set $Q$ of states, the set $F$ of final states and the initial state are the same for both the SAFA $M$ and the $k$-bag CCA $A$. Each bag of the $k$-bag CCA $A$ represents a set of the SAFA $M$. When the SAFA $M$ reads a data value $d$ and checks for $p(h_m)$, it is equivalent to CCA $A$ implementing $c_m=(>,0)$, which checks if $\beta(d)>0$ in $m^{th}$ bag. The value $\beta(d)>0$ denotes that the $k$-bag CCA $A$ has seen the data value $d$ before, and it has been recorded in the $m^{th}$ bag. When the SAFA $M$ inserts a data value $d$ in set $h_\ell$, then the CCA $A$ increments $\beta(d)$ of the $\ell^{th}$ bag by $1$. This reflects that CCA $A$ has encountered data value $d$, which is recorded in the $\ell^{th}$ bag.
    \item The transitions in the transition relation $\delta$ of SAFA $M$ are mapped to transitions in the transition relation $\Delta$ of the $k$-bag CCA $A$ as follows:
    \begin{itemize}
        \item For every transition of the form $(q_i,a,p(h_m),-,q_j)$ in the transition relation $\delta$ of SAFA $M$ which takes the SAFA $M$ from state $q_i$ to state $q_j$ where $a\in \Sigma$, $q_i,q_j\in Q$ and $h_m\in H$, we have the transition $(q_i,a,c_1,\ldots ,c_k,instr_1,\ldots ,instr_k,q_j)$ where $c_m$ is $(>,0)$ and all other $c_n$ are $(\geq,0)$, that is, where $n \neq m$, $instr_r=(\uparrow^+,0)$ for all $1\leq r \leq k$ in the transition relation $\Delta$ of the $k$-bag CCA $A$. 
        
         \item For every transition of the form $(q_i,a,!p(h_m),-,q_j)$ in the transition relation $\delta$ of SAFA $M$ which takes the SAFA $M$ from state $q_i$ to state $q_j$ where $a\in \Sigma$, $q_i,q_j\in Q$ and $h_m\in H$, we have the transition $(q_i,a,c_1,\ldots ,c_k,instr_1,\ldots ,instr_k)$ where $c_m$ is $(=,0)$ and all other $c_n$ are $(\geq,0)$, that is, where $n \neq m$, $instr_r=(\uparrow^+,0)$ for all $1\leq r \leq k$ in the transition relation $\Delta$ of $k$-bag CCA. Here, the CCA $A$ checks if $\beta_m(d)=0$, denoting  that the $k$-bag CCA $A$ has either not seen the data value 
         $d$ or it has not been recorded in the $m^{th}$ bag.

          \item For every transition of the form $(q_i,a,p(h_m)),\ins(h_\ell),q_j)$ in the transition relation $\delta$ of the SAFA $M$ which takes the SAFA $M$ from state $q_i$ to state $q_j$ where $a\in \Sigma$, $q_i,q_j\in Q$ and $h_m,h_\ell \in H$, we have the transition $(q_i,a,c_1,\ldots ,c_k,instr_1,\ldots ,instr_k)$ where $c_m$ is $(>,0)$ and all other $c_n$ are $(\geq,0)$, that is, where $n \neq m$, $instr_\ell=(\uparrow^+, 1)$ and all other $instr_r=(\uparrow^+,0)$ for all $r\neq \ell$ in the transition relation $\Delta$ of the $k$-bag CCA. Similar to SAFA $M$ inserting the data value $d$ in the set $h_\ell$, the $k$-bag CCA records the data value by incrementing $\beta_\ell(d)$ by $1$, while all the values for all other bags remain unchanged.

         \item For every transition of the form $(q_i,a,!p(h_m)),\ins(h_\ell),q_j)$ in the transition relation $\delta$ of SAFA $M$ which takes the SAFA $M$ from state $q_i$ to state $q_j$ where $a\in \Sigma$, $q_i,q_j\in Q$ and $h_m,h_\ell\in H$, we have the transition $(q_i,a,c_1,\ldots ,c_k,instr_1,\ldots ,instr_k)$ where $c_m$ is $(=,0)$ and all other $c_n$ are $(\geq,0)$, that is, where $n \neq m$, $instr_\ell=[\uparrow^+, 1]$ and all other $instr_r=(\uparrow^+,0)$ for all $r\neq \ell$ in the transition relation $\Delta$ of the $k$-bag CCA. Similar to SAFA $M$ inserting the data value $d$ in the set $h_\ell$, the $k$-bag CCA records the data value by incrementing $\beta_\ell(d)$ by $1$.
    \end{itemize}
\end{itemize}

\noindent

It is easy to see that the $k$-bag CCA $A$ accepts a data word $w$ if and only if it is accepted by the SAFA $M$.

We know from~\cite{13} that for every $k$-bag CCA, there exists a one bag CCA, which accepts the same language. Therefore, for every SAFA accepting a language $L$, there exists a CCA which accepts the same language. Now, we show that CCA are more expressive than SAFA. Recall that SAFA can accept the languages $L_{\sf fd}(a)$ and $L_{a\exists b}$ but not the language $L_{{\sf fd}(a)}\cap L_{a\exists b}$ (see Lemma~\ref{lem:intersection}). Since there exists a CCA for every language accepted by a SAFA, the languages $L_{\sf fd}(a),L_{a\exists b}\in \lcca$. Since CCA are closed under intersection~\cite{13}, we have that $L_{{\sf fd}(a)}\cap L_{a\exists b}\in \lcca$. Therefore, $\lsafa \subsetneq \lcca$.
\end{proof}

We know from~\cite{13} that $\lcca \subsetneq \lcma$ and from the above result, we get the following theorem. 

\begin{corollary}\label{corr:cma}
$\lsafa \subsetneq \lcma$.
\end{corollary}
 
We discuss the expressiveness of HRA with respect to SAFA below.

\begin{theorem}
$\lsafa \subsetneq \lhra.$
\end{theorem}

\begin{proof}
For every SAFA $M=(Q,\Sigma \times D,q_0,F,H,\delta_M)$ with $|H|=k$ accepting a language $L$, we can construct a $(k,0)$-HRA $A=(Q,\Sigma, q_0,[k],\delta_A,F)$, which accepts the same language $L$ in the following manner. The set $Q$ of states, the set $F$ of final states and the initial state are the same for both the SAFA $M$ and the $(k,0)$-HRA $A$. 
    Let $[k]$ be the sets of the HRA $A$.
    Each set $m \in [k]$ of the $(k,0)$-HRA $A$ corresponds to the set $h_m$ of the SAFA $M$.     
    The sets of the HRA $A$ are initialized to empty sets. 
Let $\mathcal{X}_m=\{S\subseteq [k]|m \in S\}$ be the set of all subsets of $[k]$ which contains the set $m$. Similarly, $\mathcal{X}_{!m}=\{S\subseteq [k]|m \notin S\}$ be the set of all subsets of $[k]$ which does not contain the set.
    Now, we show how the transitions of SAFA $M$ are simulated by the $(k,0)$-HRA $A$. Consider a data item $(a,d)$. We have the following cases:
    \begin{enumerate}
        
    \item Let $M$ choose a transition $(q_i,a,p(h_m),-,q_j)$ while reading it. This can be simulated in $A$ by using the transitions $(q_i,a,X,X,q_j)$, for all $X$ in $\mathcal{X}_m$, in $A$. The above-stated transitions are added because the HRA has transitions from one state to another only if the data value $d$ it is currently reading is present in exactly those sets that appear in $X$. 
    
    \item Now let us consider the case where $M$ chooses a transition of the form \\ $(q_i,a,!p(h_m),-,q_j)$. This is simulated in $A$ by transitions of the form $(q_i,a,X,X,q_j)$, for all $X$ in $\mathcal{X}_{!m}$, in $A$. 
    
    \item Transitions with insert operations in $M$ of the form $(q_i,a,p(h_m)), \ins(h_\ell),\\q_j)$ and $(q_i,a,!p(h_m)),\ins(h_\ell),q_j)$ are simulated in $A$ by transitions $(q_i,a,\\X,X\cup \{\ell\},q_j)$, for all $X$ in $\mathcal{X}_m$ and $(q_i,a,X,X\cup \{\ell\},q_j)$, for all $X$ in $\mathcal{X}_{!m}$, respectively. Here $X\cup \{\ell\}$ ensures that the data value $d$ currently read is inserted in all the sets in $X$ and also in set $\ell$.
    \end{enumerate}
     Thus if a data word is accepted by the SAFA $M$, then it is also accepted by the HRA $A$. Now, we prove the other direction, that is, if a data word is rejected by the SAFA $M$, then it is also rejected by HRA $A$.

    It should be noted that the status of the sets in HRA $A$ and SAFA $M$ is the same. Some data value is added to a set $m$ in HRA $A$ only if it is inserted into a set $h_m$ in SAFA $M$. The SAFA $M$ rejects a data word $w$, if all the parallel non-deterministic executions of SAFA $M$ on the data word $w$ ends up with one of the following two conditions:
    \begin{enumerate}
        \item The SAFA $M$ completely reads the input data word but ends in a non-accepting state.
        \item The SAFA $M$ at some position of the input data word cannot find a transition that can be applied to the data item at that position, thus fails to read the complete data word.
    \end{enumerate}
       Observe, that the nondeterministic transitions $(q_i,a,X,X,q_j)$, for all $X$ in $X_{!m}$ associated with $(k,0)$-HRA $A$ simulating a transition $(q_i,a,!p(h_m),-,q_j)$ of the SAFA $M$ are mutually exclusive i.e. depending on the status of the sets of the HRA $A$ and SAFA $M$ only one of the non-deterministic choices of the HRA $A$ for the transition $(q_i,a,!p(h_m),-,q_j)$ of $M$ can be executed.  This is, true for all other SAFA transitions $(q_i,a,p(h_m),-,q_j)$, $(q_i,a,p(h_m)),\ins(h_\ell),q_j)$ and $(q_i,a,!p(h_m)),\ins(h_\ell),q_j)$ also. Thus, let us consider a data word $w$ which the SAFA $M$ rejects, the HRA $A$ follows the same parallel paths of execution as the nondeterministic SAFA $M$ and rejects in a similar manner as the SAFA $M$. Thus, if the SAFA $M$ rejects or accepts a data word $w$, HRA $A$ follows the same sequence of transitions to accept or reject the word $w$. Moreover, history-register automata accept languages not accepted by any CMA~\cite{grigore2016history} and $\lsafa \subsetneq \lcma$ (see Corollary~\ref{corr:cma}).
    Hence the result.
\end{proof}

Next, we discuss the expressiveness of SDA with respect to SAFA.

\begin{lemma}\label{lem:SDA}
    The data language $L$ which comprises of all data words which have exactly two different data values cannot be accepted by any SDA.
\end{lemma}

\begin{proof}
    The proof is by contradiction. Let us suppose there exists an SDA $A$ that accepts $L$ with $B=\{b_1,b_2,\dots, b_n$\}, $\Sigma=\{a\}$ and a transducer $T$. Now we consider data words of length three in $L$ which can be infinitely many since the data word domain is infinite. The number of possible transitions for transducer $T$ at each position of the input word is at most $|\Sigma|\times |\psi|\times |B|\times |Q_T|$, where $Q_T$ is the state space of the transducer. The data words in $L$ of length three can assume any possible value from $\mathbb{N}$ at each position. Therefore, the transducer $T$ must have transitions and outputs defined for all the data values in $\mathbb{N}$ for all possible positions in the words. The Presburger formula at each position needs to be generic to accept any possible data value from $\mathbb{N}$.
    
    Consider data words of length three of the form $(a,d_i)(a,d_k)(a,d_j)$ in $L$ where $d_i, d_j, d_k\in \mathbb{N}$ and either $d_i=d_j\neq d_k$ or $d_i=d_k\neq d_j$ or  $d_k=d_j\neq d_i$. There are infinitely many such data words as $D$ is infinite. The SDA $A$, in order to accept these data words, needs the transducer $T$ to have a valid transition sequence for all these data words. A transition sequence is valid if it belongs to $L_T$. Recall, the language $L_T \subseteq (\Sigma \times \psi \times B)^*$ is a regular language. These infinitely many data words of length three in $L$ can begin with any data value in $D$. The number of possible transitions of the transducer $T$ is finite. Therefore, there is a transition $(a,\psi_\ell,b_i )$ of the transducer $T$, which begins a sequence of valid transitions and consumes an infinite set $D_i\subseteq D$ of data values. We consider only those data words of length three in $L$, which have data values from $D_i$ in its first position. Among these, we consider only those data words in $L$, which have data values from $D_i$ in the second position. There are infinitely many such data words in $L$ as $D_i$ is an infinite set. Therefore, there is a transition $(a,\psi_m,b_j)$ of the transducer $T$ that consumes an infinite set $D_x\subseteq D$ of data values such that, $D_x\cap D_i=D_j$ and $D_j$ is also an infinite set of data values in the second position of the data words. Similarly, we consider data words in $L$, where the first position has data values from $D_i$, the second position has data values from $D_j$. Among these we consider only those data words which have data values from $D_j$ in the third position. There are infinitely many such data words in $L$, since $D_i, D_j$ are infinite sets. Therefore, there is a transition $(a,\psi_r,b_k)$ of the transducer $T$ that consumes an infinite set $D_y\subseteq D$ of data values such that, $D_y\cap D_j=D_k$ and $D_k$ is also an infinite set of data values in the third position of the data words. All these words are in $L$, therefore $(a,\psi_\ell,b_i )(a,\psi_m,b_j )(a,\psi_r,b_k )\in L_T$. We can always find three distinct data values $d_1,d_2,d_3\in D_i\cap D_j\cap D_k=D_k$, since $D_k$ is an infinite set. The data words formed with various combinations of these data values, such as $w_1=(a,d_1)(a,d_2)(a,d_2), w_2=(a,d_2)(a,d_2)(a,d_3), w_3=(a,d_1)(a,d_2)(a,d_1), w_4=(a,d_1)(a,d_2)(a,d_3)$ have the same transducer output $b_i,b_j,b_k$ because the transducer has same sequence of transitions $(a,\psi_\ell,b_i )(a,\psi_m,b_j )(a,\psi_r,b_k )$ for each of these data words and the sequence of transitions $(a,\psi_\ell,b_i )(a,\psi_m,b_j )(a,\\\psi_r,b_k )\in L_T$.

    We can assume w.l.o.g that $b_i=b_1, b_j=b_2$ and $b_k=b_3$. Let us consider a data word $w_1=(a,d_1)(a,d_2)(a,d_2)\in  L$. The output of transducer $T$ w.r.t data word $w_1$ is $b_1b_2b_3$. For SDA $A$ to be able to accept the data word $w_1$, the Parikh image $(1,0,0,0^{n-3})(0,1,1,0^{n-3})$ corresponding to data value $d_1$, the Parikh image $(0,1,1,0^{n-3})(1,0,0,0^{n-3})$ corresponding to data value $d_2$, and the Parikh image $(0^n)(1,1,1,0^{n-3})$ for all other data values not present in $w_1$ must be in set $S$. Recall that $n$ is the number of output symbols of the transducer $T$ and $S$ is a semilinear set over $\mathbb{N}^B\times \mathbb{N}^B$. 
    
    Consider the data word $w_2=(a,d_2)(a,d_2)(a,d_3)\in L$. The output of transducer $T$ for $w_2$ is the same as $w_1$. 
   For SDA $A$ to be able to accept the data word $w_2$, the Parikh image $(1,1,0,0^{n-3})(0,0,1,0^{n-3})$ corresponding to the data value $d_2$, the Parikh image $(0,0,1,0^{n-3})(1,1,0,0^{n-3})$ corresponding to the data value $d_3$, and the Parikh image $(0^n)(1,1,1,0^{n-3})$ for all other data values not present in $w_2$ must be in set $S$. 
    
    Similarly, consider the data word $w_3=(a,d_1)(a,d_2)(a,d_1)\in L$. The output of transducer $T$ for $w_3$ is also the same as $w_1$. 
    For SDA $A$ to be able to accept the data word $w_3$, the Parikh image $(1,0,1,0^{n-3})(0,1,0,0^{n-3})$ corresponding to the data value $d_1$, the Parikh image $(0,1,0,0^{n-3})(1,0,1,0^{n-3})$ corresponding to the data value $d_2$, and the Parikh image $(0^n)(1,1,1,0^{n-3})$ for all other data values not present in $w_3$ must be in set $S$. Therefore, to accept data words $w_1,w_2,w_3$, the set $S$ of SDA $A$ must contain the above Parikh images.
    
    Now, let us look at the data word $w_4=(a,d_1)(a,d_2)(a,d_3)$. The output of transducer $T$ for $w_4$ is also the same as $w_1$. 
    The Parikh image corresponding to data value $d_1$ for the data word $w_4$ is $(1,0,0,0^{n-3})(0,1,1,0^{n-3})$ which is the same as the Parikh image of $d_1$ in the data word $w_1\in L$. Thus, it is present in $S$. Again, the Parikh image corresponding to data value $d_2$ for the data word $w_4$ is $(0,1,0,0^{n-3})(1,0,1,0^{n-3})$ which is the same as the the Parikh image of $d_2$ in the data word $w_3\in L$. Thus it is present in $S$. Similarly, the Parikh image corresponding to data value $d_3$ for the data word $w_4$ is $(0,0,1,0^{n-3})(1,1,0,0^{n-3})$ which is the same as the Parikh image of $d_3$ in the data word $w_2\in L$. Thus, it is present in $S$. Moreover, the Parikh image $(0^n)(1,1,1,0^{n-3})$ corresponding to all other data values not in $w_4$ is also present in $S$. Thus, the data word $w_4\notin L$ is accepted by SDA $A$, thereby leading to a contradiction.
\end{proof}

The data language $L$ which comprises of all data words which have exactly two different data values can be accepted by a SAFA $M$ with one set, where we store the first two distinct values we come across in the data word in a set and if we come across some other new data value in the data word we reject the input data word, we also reject the input data word if we come across only one distinct value, otherwise we accept the input data word. We thus get the following as a corollary of the above result.

\begin{corollary}
    $L_{SDA}$ and $L_{SAFA}$ are incomparable.
\end{corollary}

\section{Conclusion} \label{sec7}
In this paper, we introduce set augmented finite automata that use a finite set of containers with sets of data values to accept data languages. 
We have shown examples of several data languages that our model can accept. 
We compare the language acceptance capabilities of SAFA with $k$-register automata, CCA, CMA, SDA, and history-register automata. 
The computational power and low complexity of nonemptiness and membership of SAFA make it a useful tool for modeling and analysis of data languages. 
We show that the universality problem is undecidable already for singleton SAFA similar to guess register automata.
We also study the deterministic variant of SAFA, which is closed under complementation and has decidable universality.
Our model is robust enough and can also be extended to infinite words. 
This model opens up some interesting avenues for future research.
We would like to explore augmentations of SAFA with Boolean combinations of tests and the feature of updating multiple sets simultaneously. 
Updating multiple sets may lead to a more well-behaved (closed under various set operations) model with respect to closure properties. 

\subsubsection*{Acknowledgements}
We thank Amaldev Manuel for providing useful comments on a preliminary version of this paper.
We thank Yoav Danieli for pointing us to the model of register automata with guessing by Czerwiński et al.~\cite{czerwinski2021new}.

\bibliographystyle{elsarticle-num-names} 
\bibliography{refs}





\end{document}